\documentclass[prd,showpacs,notitlepage]{revtex4-1}
\usepackage{amsmath}
\usepackage{epsfig}
\usepackage{bm}
\usepackage{amssymb}
\usepackage[bookmarks=false]{hyperref}
\usepackage{graphicx}
\usepackage{color}

\newcommand{\e}{\epsilon}

\begin{document}

\title{Event-by-event distributions of azimuthal asymmetries in
ultrarelativistic heavy-ion collisions}
\author{H.~Niemi${}^{a}$, G.~S.~Denicol${}^{b,c}$, H.~Holopainen${}^{d}$,
and P.~Huovinen${}^{c,d}$}
\affiliation{$^{a}$Department of Physics, P.O.Box 35, FI-40014 University of
Jyv\"askyl\"a, Finland}
\affiliation{$^{b}$Department of Physics, McGill University, 3600 University Street,
Montreal, Quebec, H3A\,2T8, Canada}
\affiliation{$^{c}$Institut f\"ur Theoretische Physik, Johann Wolfgang
Goethe-Universit\"at, Max-von-Laue-Str.\ 1, D-60438 Frankfurt am Main,
Germany}
\affiliation{$^{d}$Frankfurt Institute for Advanced Studies, Ruth-Moufang-Str.\ 1,
D-60438 Frankfurt am Main, Germany}

\begin{abstract}
Relativistic dissipative fluid dynamics is a common tool to describe the
space-time evolution of the strongly interacting matter created in
ultrarelativistic heavy-ion collisions. For a proper comparison to
experimental data, fluid-dynamical calculations have to be performed on an
event-by-event basis. Therefore, fluid dynamics should be able to reproduce,
not only the event-averaged momentum anisotropies, $\left\langle
v_{n}\right\rangle $, but also their distributions. In this paper, we
investigate the event-by-event distributions of the initial-state
and momentum anisotropies $\epsilon_n$ and $v_n$, and their
correlations. We demonstrate that the event-by-event distributions of
relative $v_n$ fluctuations are almost equal to the event-by-event
distributions of corresponding $\epsilon_n$ fluctuations, allowing
experimental determination of the relative anisotropy
fluctuations of the initial state.
Furthermore, the correlation $c(v_2,v_4)$ turns out to be sensitive to the
viscosity of the fluid providing an additional constraint to the properties
of the strongly interacting matter.
\end{abstract}

\maketitle

\section{Introduction}

Relativistic dissipative fluid dynamics is the most widely employed model 
to describe the space-time evolution of the quark-gluon plasma (QGP) created
in ultrarelativistic heavy-ion collisions. It was the success of
fluid-dynamical models in describing the large azimuthal momentum
anisotropies observed in heavy-ion collisions that led to our current
picture of the QGP, as a strongly interacting fluid with one of the smallest
shear viscosity to entropy density ratios, $\eta/s$, ever observed \cite%
{Gyulassy:2004zy}.

The azimuthal momentum anisotropy is characterized in terms of the
coefficients $v_{n}$ of the Fourier expansion of the single particle
azimuthal distribution: 
\begin{eqnarray}
\frac{\mathrm{d} N}{\mathrm{d} y\mathrm{d}\phi} & = & \frac{\mathrm{d} N}{%
\mathrm{d} y}\left[ 1+2v_{1}\cos \left( \phi -\psi_1 \right) +2v_{2}\cos %
\left[ 2\left( \phi -\psi _{2}\right) \right] +\ldots \right] ,  \notag \\
v_{n} & = & \frac{\int \mathrm{d}\phi \cos \left[ n\left( \phi -\psi
_{n}\right) \right] \frac{\mathrm{d} N}{\mathrm{d} y\mathrm{d}\phi }} {\int 
\mathrm{d}\phi \frac{\mathrm{d} N}{\mathrm{d} y\mathrm{d}\phi }}
=\left\langle \cos \left[ n\left( \phi -\psi _{n}\right) \right]
\right\rangle,  \label{v_n}
\end{eqnarray}%
where $\psi_{n}$ is the event-plane angle, $\psi_{n}  = (1/n) \arctan
\left(\left\langle p_T \sin n\phi \right\rangle/  \left\langle p_T \cos
n\phi \right\rangle\right)$, and $\phi $ is the transverse momentum
azimuthal angle. One of the main features of the fluid-dynamical description
of the expansion is that the anisotropy originates from the
azimuthal anisotropy of the initial density profile. In the literature this
initial anisotropy is quantified in terms of 
coefficients $\epsilon_{m,n}$: 
\begin{equation}
  \epsilon_{m,n} = -\frac{\int \mathrm{d} x \mathrm{d} y \; r^{m} \cos \left[
  n\left( \phi -\Psi_{m,n}\right) \right] \varepsilon \left(
  x,y,\tau_{0}\right) } {\int \mathrm{d} x \mathrm{d} y \; r^{m} \varepsilon
  \left( x,y,\tau _{0}\right) },  \label{e_n}
\end{equation}%
where $\varepsilon$ is the energy density, $r^{2}=x^{2}+y^{2}$, $\phi$ is
now the spatial azimuthal angle, and $\Psi_{m,n}$
is the participant angle, defined as
\begin{equation}
  \Psi_{m,n} = \frac{1}{n} \arctan 
  \frac{\int \mathrm{d}x \mathrm{d}y \; r^m \sin \left( n\phi \right) \varepsilon\left( x,y,\tau _{0}\right) } 
  { \int \mathrm{d}x \mathrm{d}y \; r^m \cos \left( n\phi \right) \varepsilon\left( x,y,\tau _{0}\right)}  + \pi/n.
\end{equation}
In the following we will concentrate on the {anisotropies $\epsilon_{2,n}$,
and use a shorthand $\epsilon_n \equiv \epsilon_{2,n}$.

In fluid-dynamical calculations, a linear relation between $v_{2}$ and $%
\epsilon_{2}$ was found, \emph{i.e.}, $v_{2} \propto \epsilon_2$~\cite%
{Kolb:2003dz}. The proportionality coefficient
was shown to depend, not only on the properties of the fluid
such as the equation of state and viscosity, but also on the initial
density, freeze-out temperature, and resonance content of the late hadronic
state~\cite{Kolb:2003dz,Song:2008si,Huovinen:2003fa}. 

The initial conditions in these fluid dynamical calculations were
always smooth, constructed as an average over infinitely many
individual collisions of the particular centrality. It was thought
that the use of this kind of averaged initial conditions would lead
to a good description of observables which were averaged over many
events. In other words, the anisotropy $v_2$ computed using the
event-averaged initial condition was expected to be equal to the $\langle
v_2 \rangle_\mathrm{ev}$ observed in the collisions, where $\langle \ldots
\rangle_\mathrm{ev}$ corresponds to an average over all the events of the
corresponding centrality class.

Recently, it has been realized that, in order to obtain a proper comparison
with experimental data, fluid-dynamical calculations have to be performed on
an event-by-event basis. This was first pointed out
by Kodama \textit{et al.}~\cite{Andrade:2006yh} almost ten years ago. 
However, this view only become
widely accepted years later, after the work of Alver and Roland~\cite%
{Alver:2010gr}. They showed that $\epsilon_3$ and, consequently, $v_3$
are non-zero in a single event. This is in
contrast to the traditionally used event-averaged initial conditions
in fluid-dynamical models which had zero $\epsilon_3$
and $v_3$. Furthermore, Alver and Roland demonstrated that such finite value
of $v_3$ can be observed in heavy-ion collisions. This
finding made $\left\langle v_{3}\right\rangle_{\mathrm{ev}}$ 
as important observable as
$\left\langle v_{2}\right\rangle_{\mathrm{ev}}$ for probing
the properties of the dense matter formed in heavy-ion collisions, and led
to several works studying the behavior of observables in an
event-by-event fluid-dynamical description \cite{Alles,Holopainen:2010gz}.

On the other hand, if fluid dynamics can be applied to describe individual
ultrarelativistic heavy-ion collisions, it must be able to describe 
$v_n$ in every collision, not only the average $\langle v_n\rangle_\mathrm{ev%
}$. Therefore it must be able to reproduce the distribution 
$\mathcal{P}(v_n)$ 
of $v_n$ in an ensemble of events too. To confirm the applicability of
fluid dynamics to describe the expansion stage of heavy-ion collisions, it
is thus not enough to check whether the event-averaged values of $v_{n}$
agree with the data, but one must also check whether their distributions, $%
\mathcal{P}\left( v_{n}\right) $, match what is experimentally observed.
Recently, the distributions of $v_2$, $v_3$, and $v_4$ were measured at the
LHC by the ATLAS collaboration \cite{Jia}. Also, the first fluid-dynamical
calculations of these distributions were performed by 
Gale \textit{et al.}~\cite{Schenke2}.

In this paper, we study the event-by-event probability distribution of the
Fourier coefficients $v_{n}$, $P\left( v_n\right)$, and how
they are correlated with the initial state anisotropies $\epsilon _{n}$
event-by-event. The goal of this paper is not to attempt a comparison with
experimental data, but to explore how these distributions and correlations
are affected by the fluid viscosity and initialization
of the
system. In this way, it will be possible to understand what can be learned
by measuring such event-by-event distributions.

In the following we explain our fluid dynamical model in
section~\ref{model}, and show our results in section~\ref{Results}.
Subsection \ref{Correlations} is dedicated to an analysis
of the event-by-event correlation between initial
condition and flow anisotropy, while in subsections \ref{distributions} and %
\ref{Lcorrelations} we show our results for 
probability distributions of scaled anisotropy $\delta v_n$, $P(\delta v_n)$,
and linear correlation coefficients $c(v_n,v_m)$, respectively.
In section \ref{conclusions}, 
we summarize our findings and make our conclusions.

\section{Model}

\label{model}

To generate the initial states event-by-event, we use a Monte-Carlo Glauber
model as implemented in Ref.\ \cite{Holopainen:2010gz}. In this model,
nucleons are distributed into nuclei according to Woods-Saxon distribution.
NN-correlations and finite size effects are neglected since they have a
negligible effect on the anisotropy coefficients \cite{Holopainen:2}. In an
event with a given impact parameter, nucleons from different nuclei are
assumed to collide when their transverse distance $d$ is small enough, 
\textit{i.e.}, when $d^2 < \sigma_{NN}/\pi$.

We consider two initial conditions, in which the initial entropy density, $s$%
, at $\tau _{0}=1$ fm, is evaluated as 
\begin{equation}
s\left( x,y\right) = W\sum_{i=1}^{N_{\mathrm{part, bin}}} \exp \left\{ -%
\left[ \left( x-x_{i}\right) ^{2}+\left( y-y_{i}\right) ^{2}\right] /\left(
2\sigma ^{2}\right) \right\} ,
\end{equation}%
where $x_{i}$ and $y_{i}$ are the spatial coordinates of either wounded
nucleons (initial condition sWN) or binary collisions (initial condition
sBC), given by the Monte-Carlo Glauber model. $W$ is a normalization
constant fixed to provide the observed multiplicity and $\sigma =0.8$ fm is
the spatial scale of a wounded nucleon or a binary collision. The centrality
classes are determined according to the number of binary collisions (for
initial condition sBC) or the number of participants (for initial condition
sWN). The initial fluid velocity and shear-stress tensor are set to zero and
we neglect the effects of bulk viscosity.

For the fluid-dynamical evolution, we use the model previously employed in
Ref.\ \cite{Niemi:2012ry}. We describe the time evolution of the fluid in
the central rapidity region assuming boost invariance and a zero
baryochemical potential. The equations of motion are given by the
conservation laws for energy and momentum: 
\begin{equation}
\partial_{\mu}T^{\mu \nu }=0,
\end{equation}
where $T^{\mu\nu}=\left( \varepsilon +p\right) u^{\mu}u^{\nu}-g^{\mu \nu}p
+\pi ^{\mu \nu }$, with $\varepsilon $, $p$, $u^{\mu }$, and $\pi^{\mu\nu }$
being the energy density, the thermodynamic pressure, the fluid 4-velocity,
and the shear-stress tensor, respectively. We use the lattice QCD and hadron
resonance gas based equation of state $s95p$-PCE-v1~\cite{ Huovinen:2009yb}
with chemical freeze-out at temperature $T_{chem} = 150$~MeV. The evolution
equation of the shear-stress tensor is given by transient relativistic fluid
dynamics~\cite{IS, DNMR}: 
\begin{eqnarray}
\Delta^{\mu\nu}_{\alpha \beta}\tau_\pi D\pi^{\alpha\beta} &+& \pi^{\mu\nu} =
2\eta \sigma^{\mu \nu} -\frac{4}{3} \pi^{\mu\nu} \theta -\frac{10}{7}
\Delta^{\mu\nu}_{\alpha \beta}\sigma^{ \alpha}_{\,\,\,\, \lambda} \pi^{\beta
\lambda} +\frac{74}{315\eta} \Delta^{\mu\nu}_{\alpha \beta}\pi^{
\alpha}_{\,\,\,\, \lambda} \pi^{\beta \lambda},  \label{eq:IS}
\end{eqnarray}
where $\eta $ is the shear viscosity coefficient, $D=u^{\mu }\partial _{\mu }
$ is the co-moving time derivative, $\sigma^{\mu \nu
}=\Delta^{\mu\nu}_{\alpha \beta} \partial^{ \alpha }u^{ \beta}$ is the shear
tensor, $\theta = \partial_\mu u^\mu$ is the expansion rate, and $%
\Delta^{\mu\nu}_{\alpha \beta} = \left(\Delta^\mu_\alpha \Delta^\nu_\beta
+\Delta^\nu_\alpha \Delta^\mu_\beta -
2/3\Delta^{\mu\nu}\Delta_{\alpha\beta}\right)/2$, with $\Delta^{\mu\nu} =
g^{\mu\nu} - u^\mu u^\nu$. The transport coefficients of the non-linear
terms on the right-hand side of the Eq.~\ref{eq:IS} were taken in the
massless limit, in the 14-moment approximation, and the relaxation time was
assumed to be $\tau _{\pi }=5\eta /\left( \varepsilon +P\right)$ \cite%
{DNMR,dkr}. Here, we have not included the nonlinear terms related to the
vorticity tensor. Note that the last two terms in Eq.~(\ref{eq:IS}) were not
included in our previous studies~\cite{Niemi:2012ry}. While such terms can
have a significant effect on many observables, they are not
relevant for the results discussed in this paper. We shall leave a detailed
investigation of the effect of such terms to a future work.
The equations of motion were solved numerically using the SHASTA
algorithm, whereas the evolution equations for shear stress
(Eq.~\ref{eq:IS}) were solved using simple finite differencing
scheme. For more details see Refs.~\cite{Niemi:2012ry, Harri EPJ}.

The hadron spectra are calculated with the Cooper-Frye freeze-out procedure 
\cite{Cooper:1974mv} using the decoupling temperature $T_{f}=100$ MeV, which
was shown to give reasonable agreement with both the $p_{T}$-spectrum and 
$\left\langle v_{2}\right\rangle_{\mathrm{ev}}$ for pions at RHIC when a
temperature-dependent $\eta/s$ was used, see Refs.\ \cite{Niemi:2012ry}. In
this work, we use constant 
values of viscosity, $\eta/s=0$ and $0.16$. Nevertheless, 
the $p_{T}$-spectrum and $\left\langle v_{2}\right\rangle_{%
\mathrm{ev}}$ remain close to what is actually observed at RHIC. Since our
main purpose is not the comparison to experimental observables, we
adjusted only the initial entropy density to fit the observed
multiplicity, but kept all the other parameters unchanged.
Finally, we use Israel and Stewart's 14-moment ansatz
for the dissipative correction to the local equilibrium distribution
function, 
\begin{equation}
\delta f_i = f_{0i} \frac{p_i^\mu p_i^\nu \pi_{\mu\nu}}{T^2\left(%
\varepsilon+p\right)},  \label{eq:deltaf}
\end{equation}
where $f_{0i} = \left\{\exp\left[\left(u_\mu p_i^\mu-\mu_i\right)/T\right]
\pm 1 \right\}^{-1}$ is the local equilibrium distribution function, with
the index $i$ indicating different hadron species and $p_i^\mu$ the
four-momentum of the corresponding hadron. 
After calculating the thermal spectra, we include the
contribution from all 2- and 3-particle decays of unstable
resonances up to 1.1 GeV mass.

It should be noted that because we do not generate particle ensembles at any
point we always know the direction of the event plane and the magnitude of
$v_n$ exactly. Experimentally, one measures a finite number of
particles, which smears the observed distribution of $v_n$. However, the final
experimental result for the $v_n$ distributions undergoes an unfolding
procedure that is supposed to remove such a smearing~\cite{Jia}.
Therefore, for a comparison with data, one can use the particle
distributions computed with fluid dynamics without generating an ensemble of
particles. A more detailed way would be to generate the particle ensembles
and apply the same complicated unfolding procedure used by the
experimentalist to obtain the $v_n$ distribution, 
but this procedure would be an
unnecessary complication for the purpose of this work.

\section{Results}

\label{Results}

In this work we consider Au+Au collisions at $\sqrt{s_{NN}}=200$ A GeV. All
the results shown in this paper are for positively charged pions. For each
centrality class a total of 2000 events were computed. The Fourier
coefficients and the initial-state anisotropies
were calculated according
to Eqs.~(\ref{v_n}), and (\ref{e_n}), respectively. In the following, we
consider two constant values for the shear viscosity to entropy density
ratio, $\eta /s=0$ and $0.16$.

\subsection{Correlations}

\label{Correlations}

As mentioned in the Introduction, it has been known for a long
time that the event averaged $v_2$, and the eccentricity of the
averaged initial state, $\epsilon_2$ are approximately linearly
related~\cite{Kolb:2003dz}. Similar relation has been found for
$\epsilon_3$ and the average $v_3$ but not for $\epsilon_4$
and $v_4$~\cite{Alver:2010dn,Qiu:2011iv}.
Here we study whether similar relations
hold event-by-event by evaluating the linear correlation between the
harmonics $v_n$ and $\epsilon_n$. We use the linear correlation
coefficient,
\begin{equation}
c\left( a,b\right) =\left\langle \frac{\left( a-\left\langle a\right\rangle
_{\mathrm{ev}}\right) \left( b-\left\langle b\right\rangle _{\mathrm{ev}%
}\right) }{\sigma _{a}\sigma _{b}}\right\rangle _{\mathrm{ev}},
\label{pearson}
\end{equation}%
where $\sigma_{a}$ is the standard deviation of the quantity $a$. This
correlation function is $1$ ($-1$) if $a$ and $b$ are linearly
(anti-linearly) correlated and zero in the absence of linear correlation.

\begin{figure}[!h]
\hspace{-0.5cm} \epsfysize 4.4cm %
\epsfbox{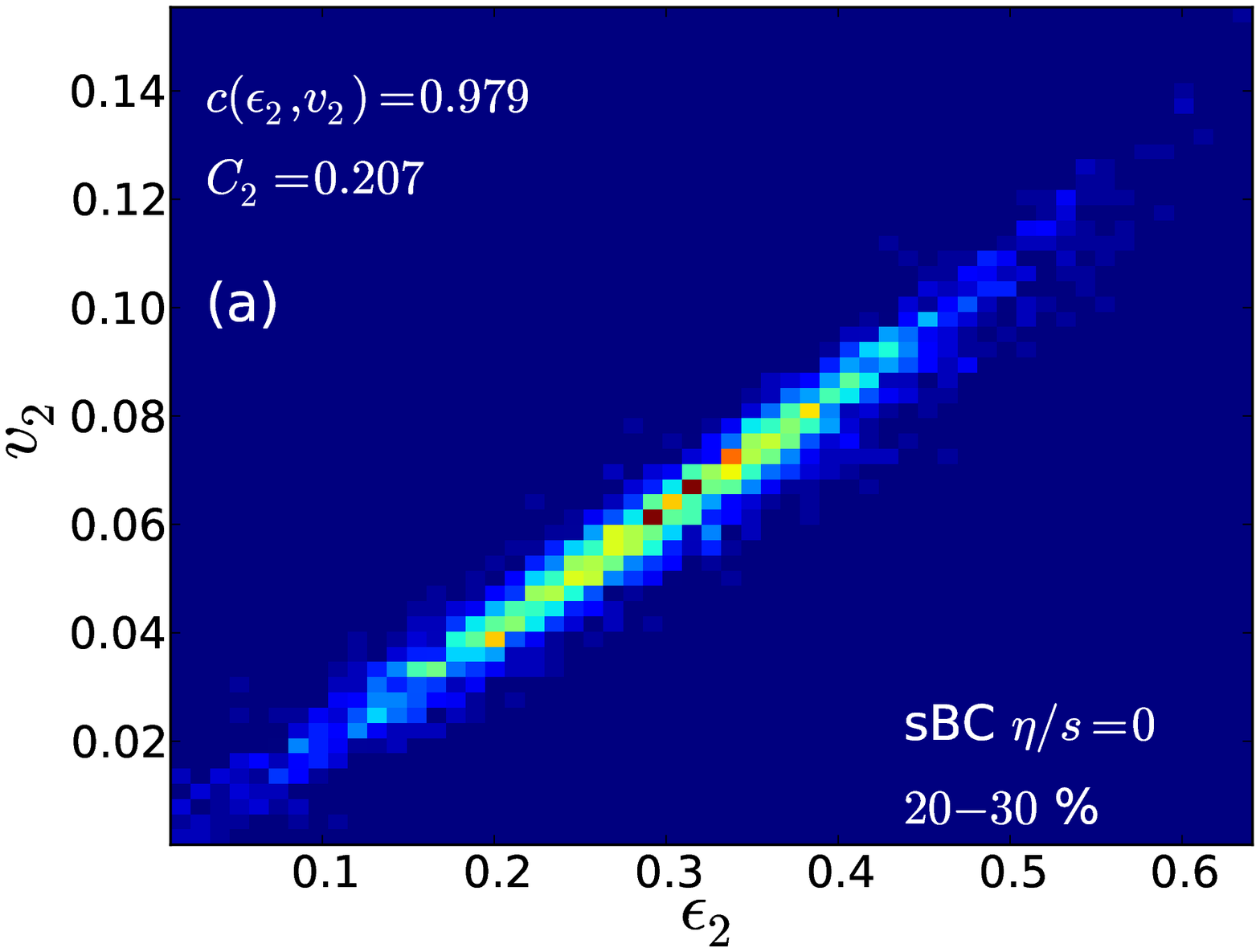} \hspace{0.0cm} %
\epsfysize 4.4cm \epsfbox{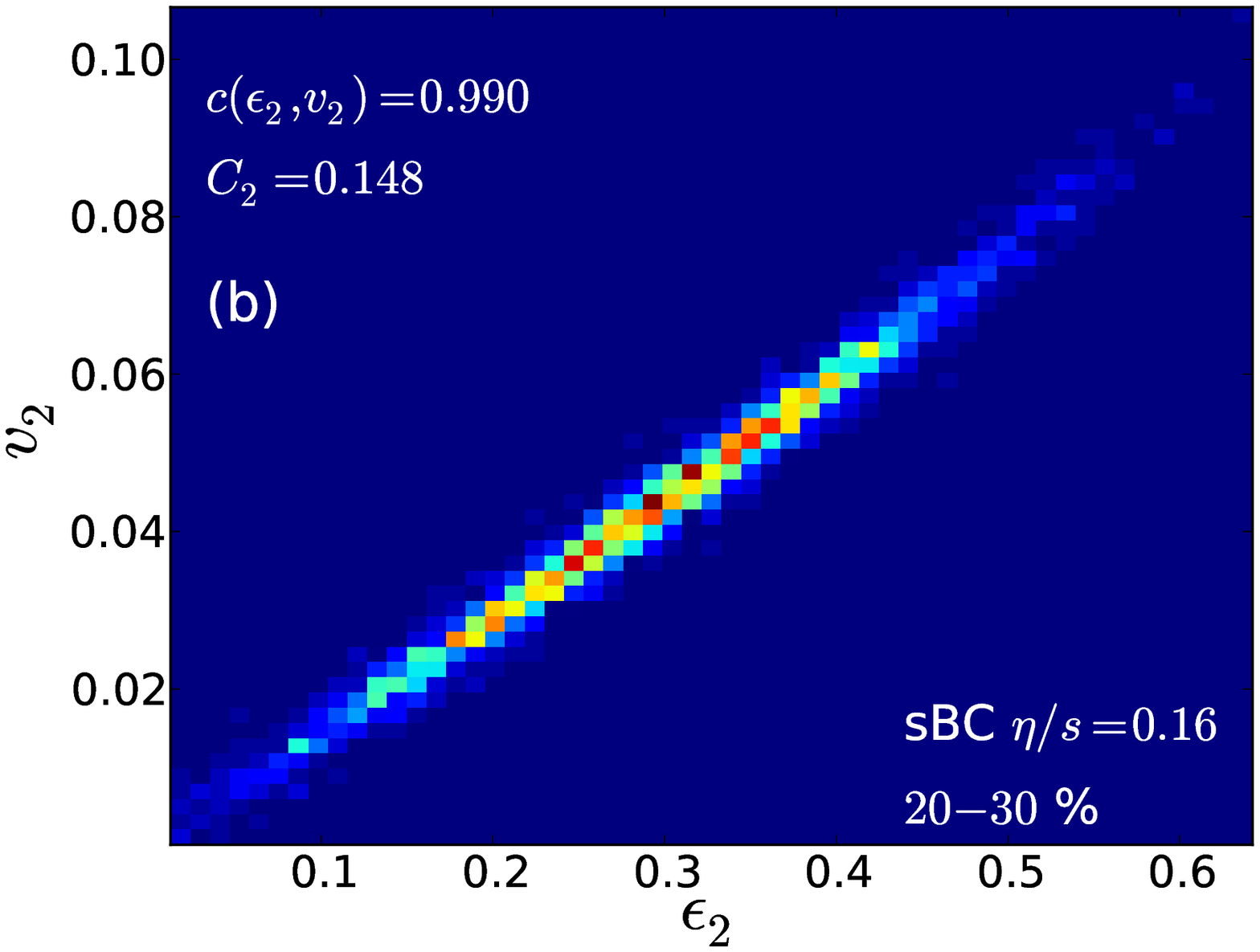} %
\epsfysize 4.4cm \epsfbox{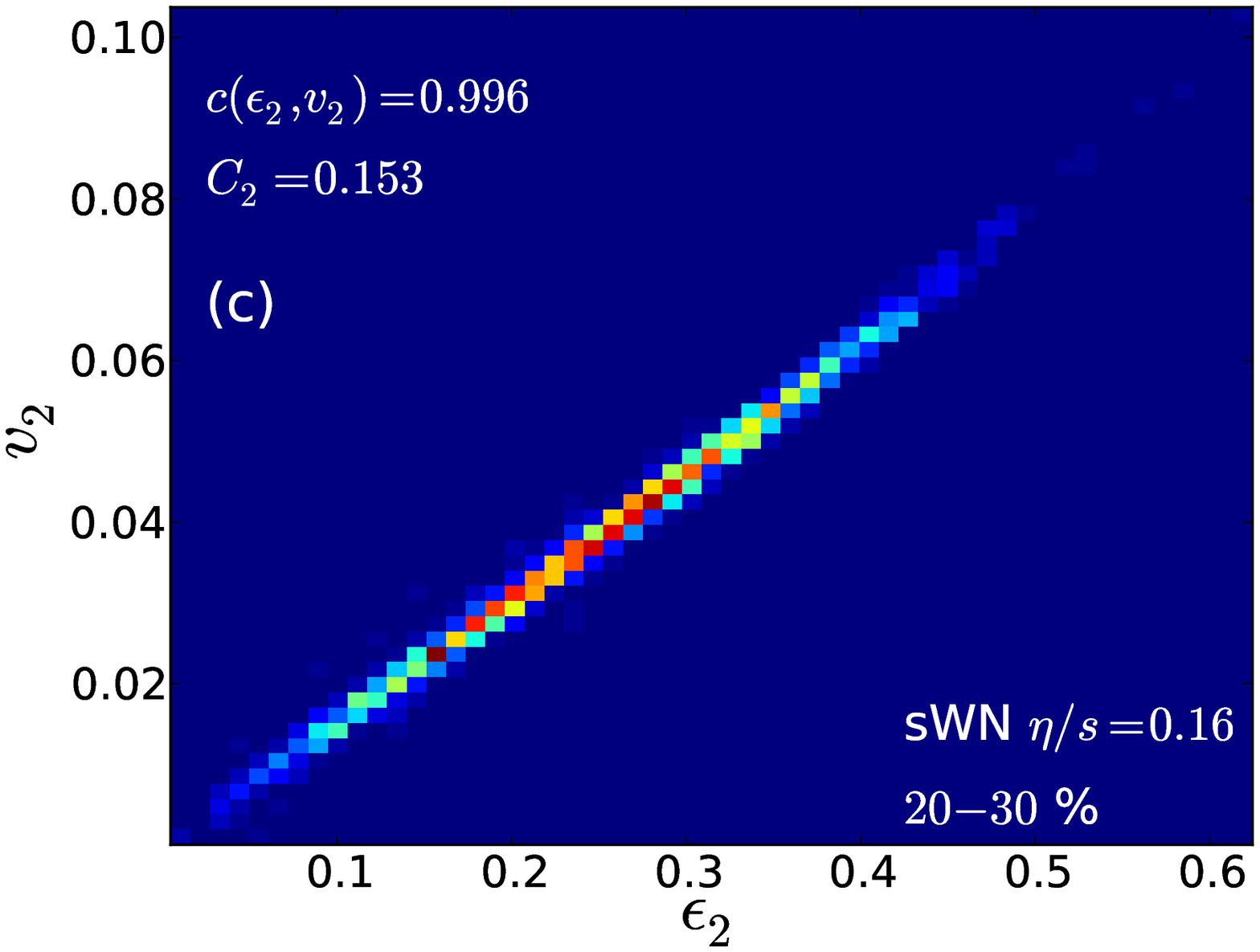} 
\caption{$\protect\epsilon_2$ and $v_2$ of pions in the $20-30$ \%
centrality class using different initializations and viscosities. a) sBC and 
$\protect\eta/s = 0$, b) sBC and $\protect\eta/s = 0.16$ and c) sWN and $%
\protect\eta/s = 0.16$.}
\label{fig:e2v2_20_30}
\end{figure}
\begin{figure}[!h]
\hspace{-0.5cm} \epsfysize 4.4cm %
\epsfbox{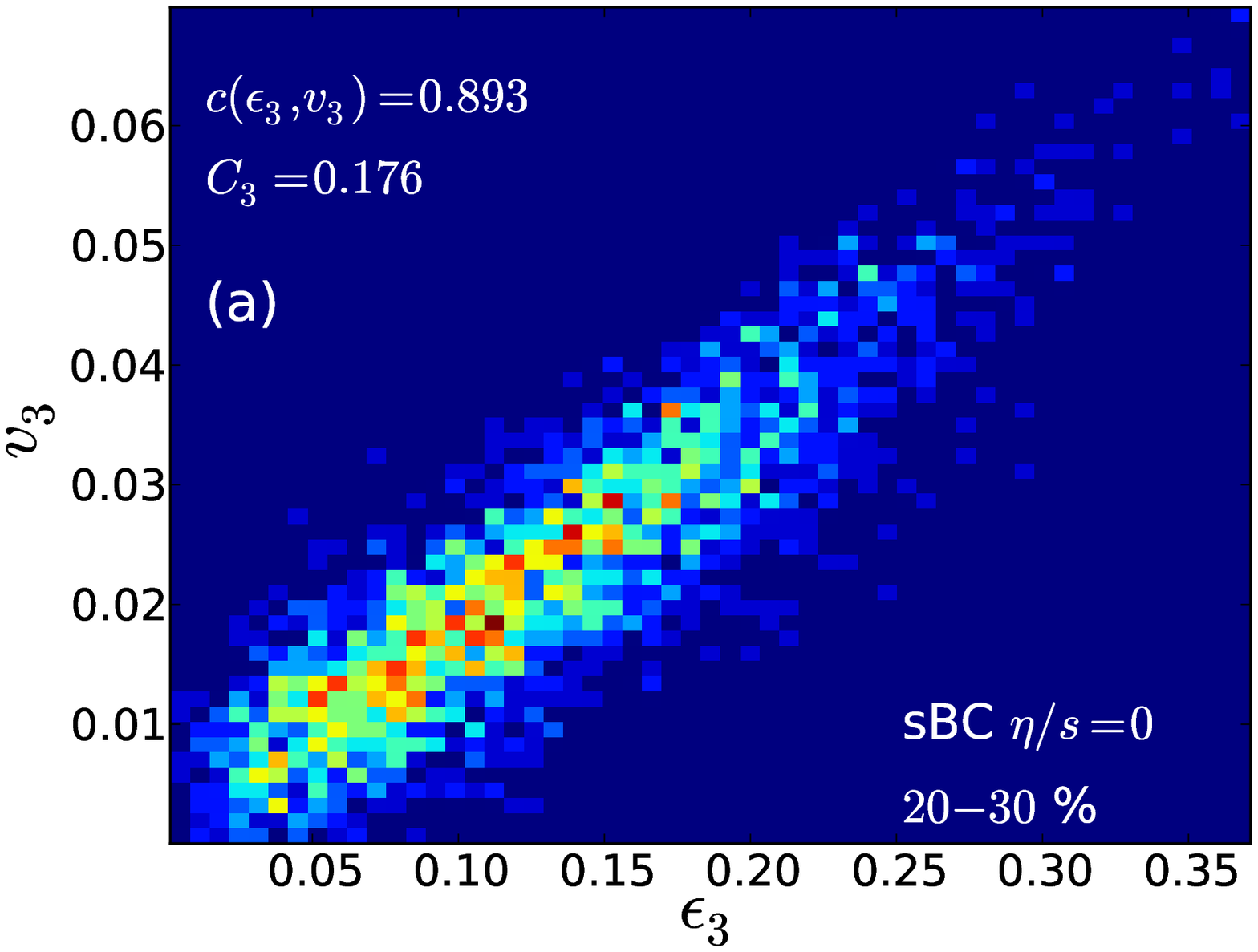} \hspace{0.0cm} %
\epsfysize 4.4cm \epsfbox{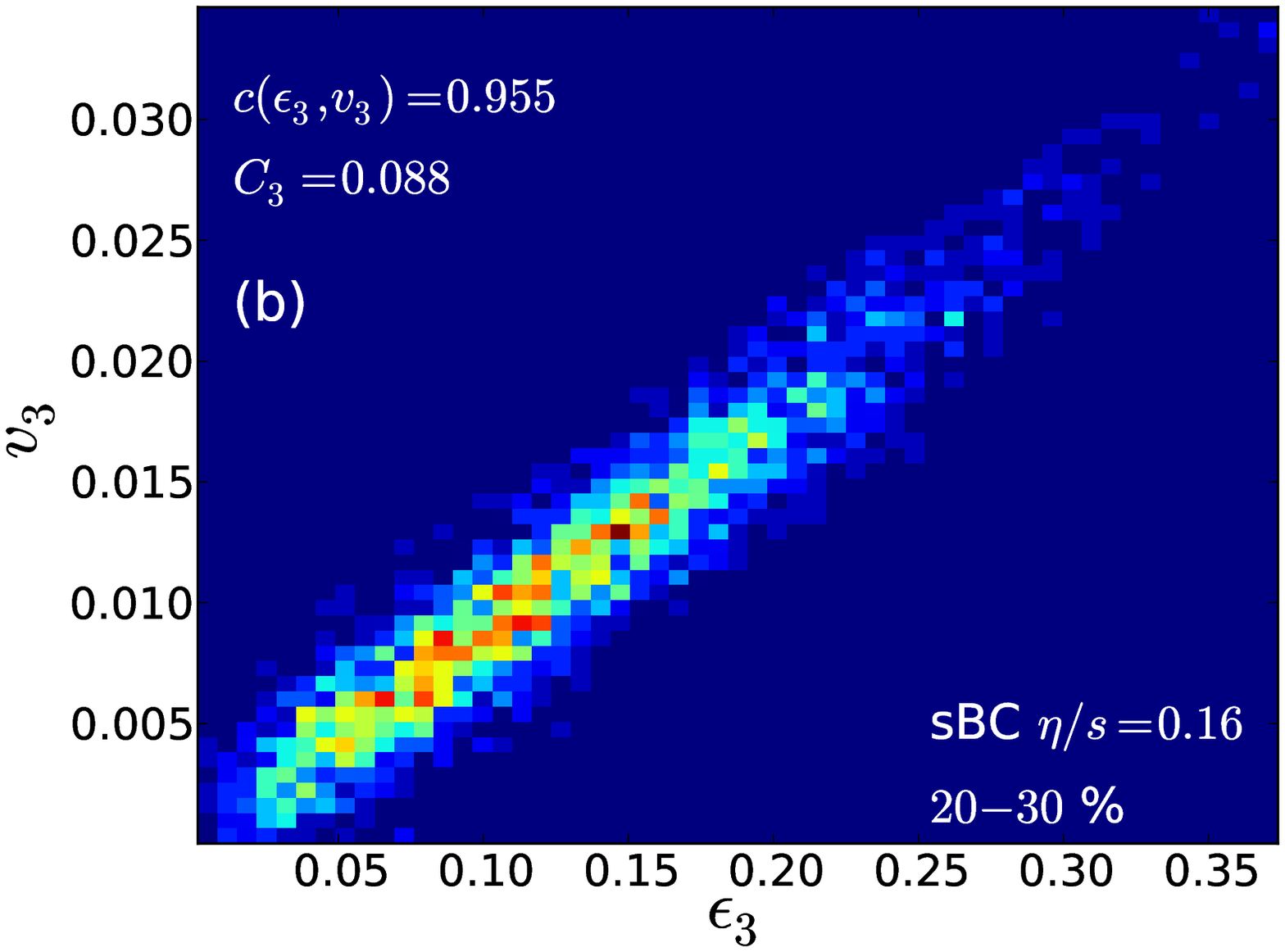} %
\epsfysize 4.4cm \epsfbox{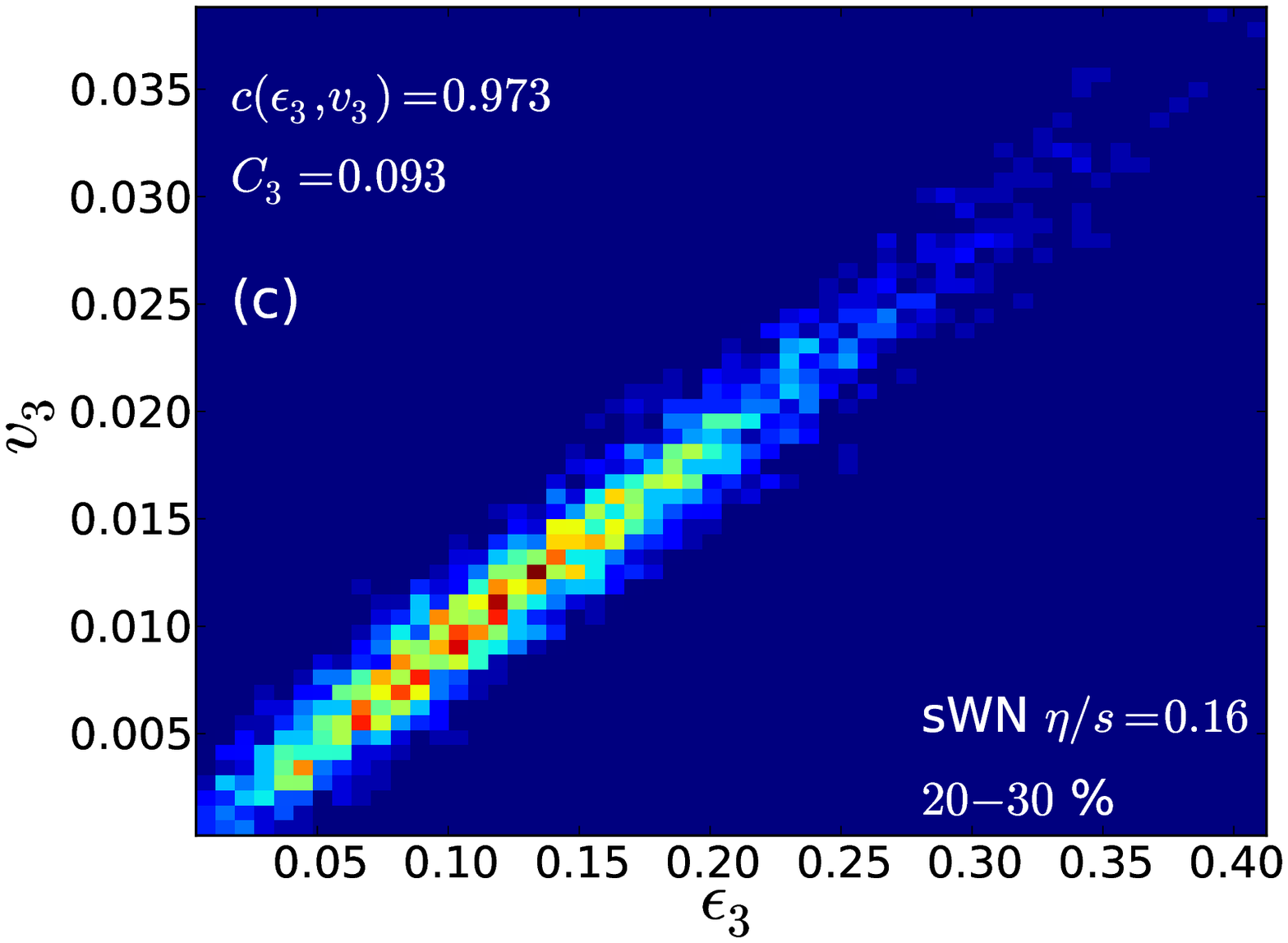} 
\caption{$\protect\epsilon_3$ and $v_3$ of pions in the $20-30$ \%
centrality class using different initializations and viscosities. a) sBC and 
$\protect\eta/s = 0$, b) sBC and $\protect\eta/s = 0.16$ and c) sWN and $%
\protect\eta/s = 0.16$.}
\label{fig:e3v3_20_30}
\end{figure}
\begin{figure}[!h]
\hspace{-0.5cm} \epsfysize 4.4cm %
\epsfbox{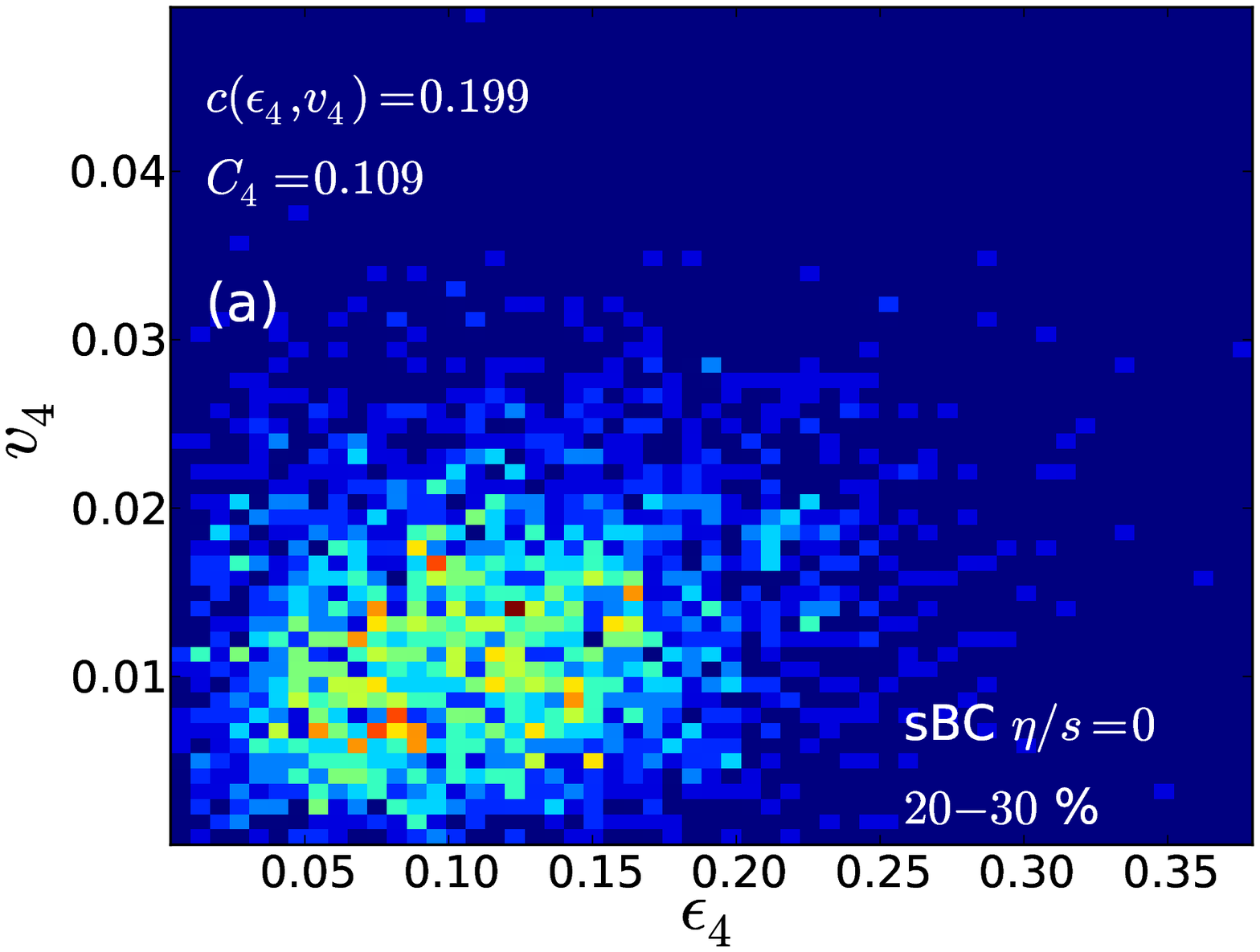} \hspace{0.0cm} %
\epsfysize 4.4cm \epsfbox{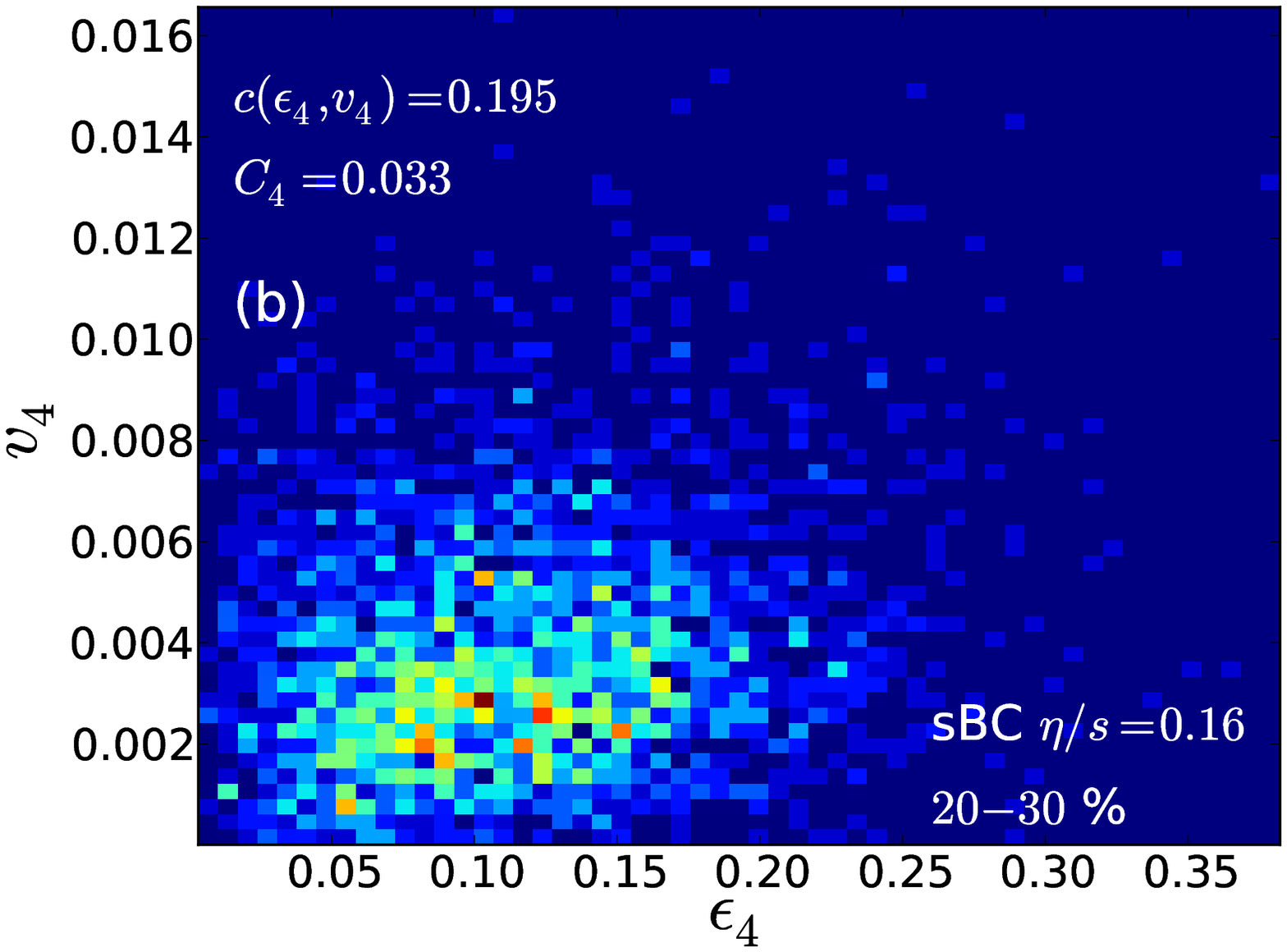} %
\epsfysize 4.4cm \epsfbox{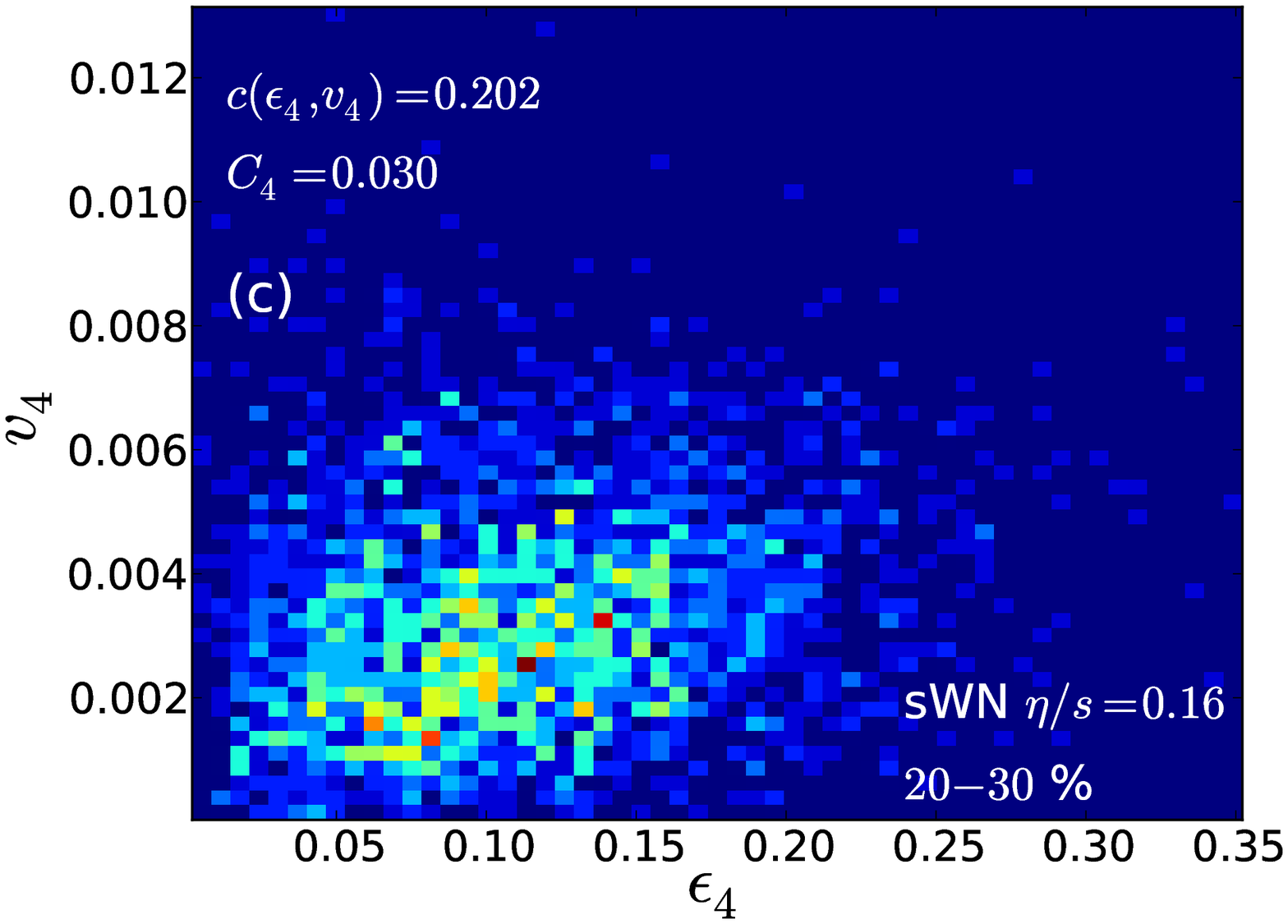} 
\caption{$\protect\epsilon_4$ and $v_4$ of pions in the $20-30$ \%
centrality class using different initializations and viscosities. a) sBC and 
$\protect\eta/s = 0$, b) sBC and $\protect\eta/s = 0.16$ and c) sWN and $%
\protect\eta/s = 0.16$.}
\label{fig:e4v4_20_30}
\end{figure}

The 2-dimensional histograms in
Figs.~\ref{fig:e2v2_20_30}--\ref{fig:e4v4_20_30} show the
correlations between $\epsilon_2$ and $v_2$, $\epsilon_3$ and $v_3$,
and $\epsilon_4$ and $v_4$, respectively, for the $20-30$ \%
centrality class. To study the effect of both viscosity and
initialization on these correlations, we show the correlations in
three different cases: (a) sBC initialization with $\eta/s = 0$, (b)
sBC initialization with $\eta/s = 0.16$, and (c) sWN initialization
with $\eta/s = 0.16$.

As can be seen in these figures, the $v_{2}$ and $v_{3}$ coefficients
display a strong linear correlation to their corresponding initial-state
coefficients for all cases considered. This is confirmed by the
values of the linear correlation coefficient $c\left( v_{2},\epsilon
_{2}\right) \sim c\left( v_{3},\epsilon _{3}\right) \sim 1$, as shown in the
Figures (top left corner). As for any two variables we can write
\begin{equation}
 v_{n}=C_{n}\epsilon _{n}+\delta _{n},
\end{equation}%
where $C_{n}=$ $\left\langle v_{n}\right\rangle _{\mathrm{ev}}/\left\langle
\epsilon _{n}\right\rangle _{\mathrm{ev}}$, and consequently, $\left\langle
\delta _{n}\right\rangle _{\mathrm{ev}}$\thinspace $=0$.
The values of $C_{n}$ are shown in Figs.~\ref{fig:e2v2_20_30} and 
\ref{fig:e3v3_20_30}. For $n=2$ a linear relation, $v_{2}=C_{2}\epsilon _{2}$, 
is approximately satisfied event-by-event
with only $\sim 10\%$ deviations
from this relation at a given $\epsilon _{2}$. On the other hand, an
event-by-event linear relation between $v_{3}$ and $\epsilon _{3}$ is not
satisfied well, with $v_{3}$ deviating $\sim 50\%$ from $%
v_{3}=C_{3}\epsilon _{3}$ at a given $\epsilon _{3}$.

In all the cases considered above, there is basically no linear
correlation between $\epsilon_4$ and $v_4$, see
Fig.~\ref{fig:e4v4_20_30}. At least one reason for this behavior is
that there is also correlation between $\epsilon^2_2$ and $v_4$, which
can be of the same order or larger than $c(\epsilon_4, v_4)$:
$c(\epsilon_2^2, v_4) = 0.40$ (sBC, $\eta/s = 0$ ), $c(\epsilon_2^2,
v_4) = 0.69$ (sBC, $\eta/s = 0.16$) and, $c(\epsilon_2^2, v_4) = 0.46$
(sWN, $\eta/s = 0.16$). This is a non-linear effect triggered by the
coupling between two different Fourier coefficients, \emph{i.e.}
$n=2$ and $n=4$, and a linear combination of these two components was
found to be a good estimator for $v_4$~\cite{Gardim:2011xv}.

As expected, the
proportionality coefficients $C_n$ are sensitive to the value of the
shear viscosity. This can be seen by comparing 
Figs.~\ref{fig:e2v2_20_30}a and \ref{fig:e2v2_20_30}b (n = 2), 
Figs.~\ref{fig:e3v3_20_30}a and \ref{fig:e3v3_20_30}b (n = 3), and
Figs.~\ref{fig:e4v4_20_30}a and \ref{fig:e4v4_20_30}b (n = 4). 
In general, the higher Fourier coefficients are
expected to be more sensitive to the viscosity~\cite{Alver:2010dn}. 
This is also the case in our
calculations, and is confirmed by comparing the relative changes
in the coefficients $C_2$, $C_3$ and $C_4$.

Note that the proportionality constants $C_n$ do not depend only on the
intrinsic properties of the fluid, but also on the initial conditions. 
Again something to be expected, since in the calculations done
using averaged initial conditions, the precise value of the
proportionality depended on many details as discussed in the
Introduction.

\begin{figure}[!h]
\hspace{-0.5cm} \epsfysize 4.4cm %
\epsfbox{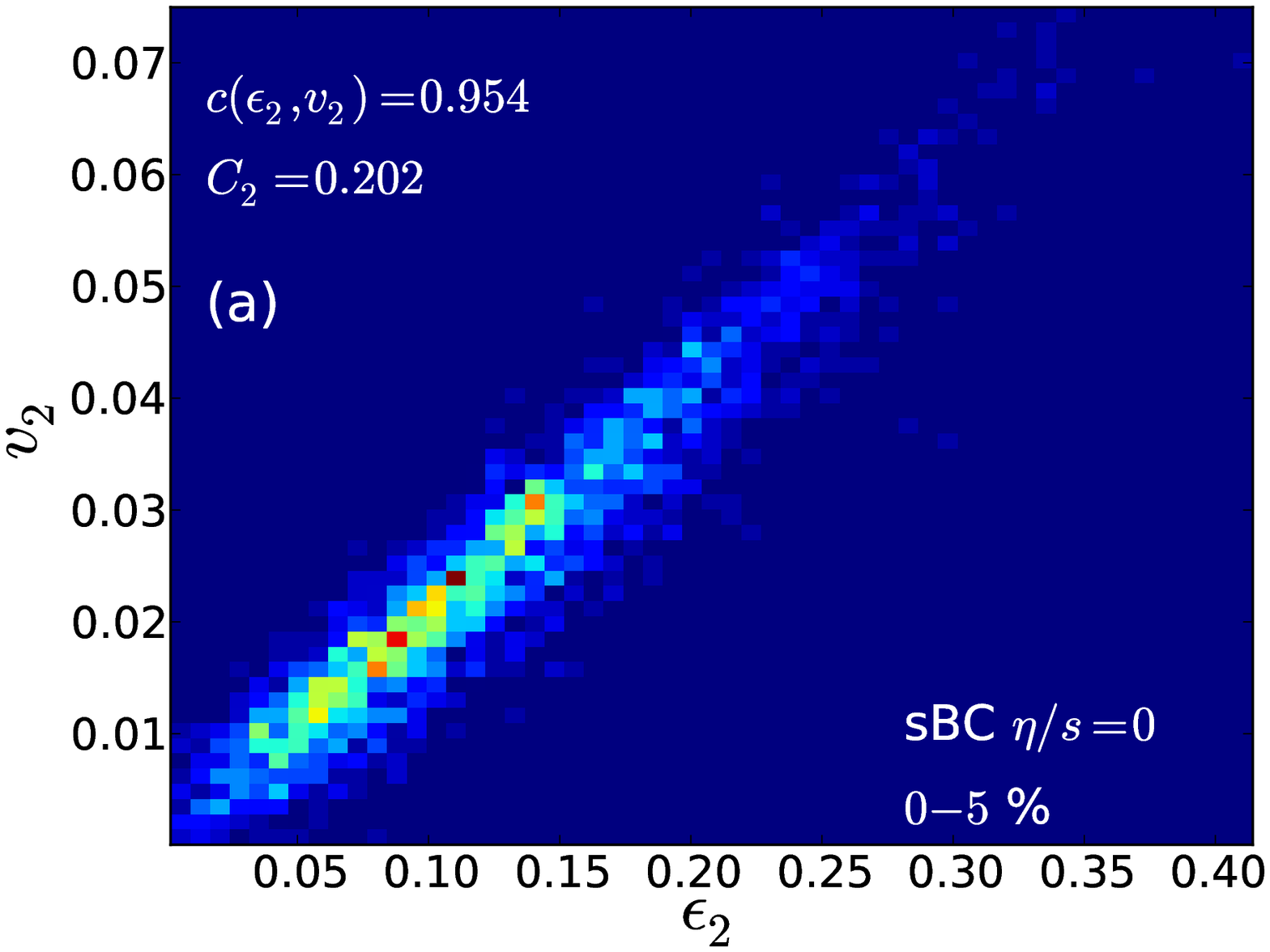} \hspace{0.0cm} \epsfysize %
4.4cm \epsfbox{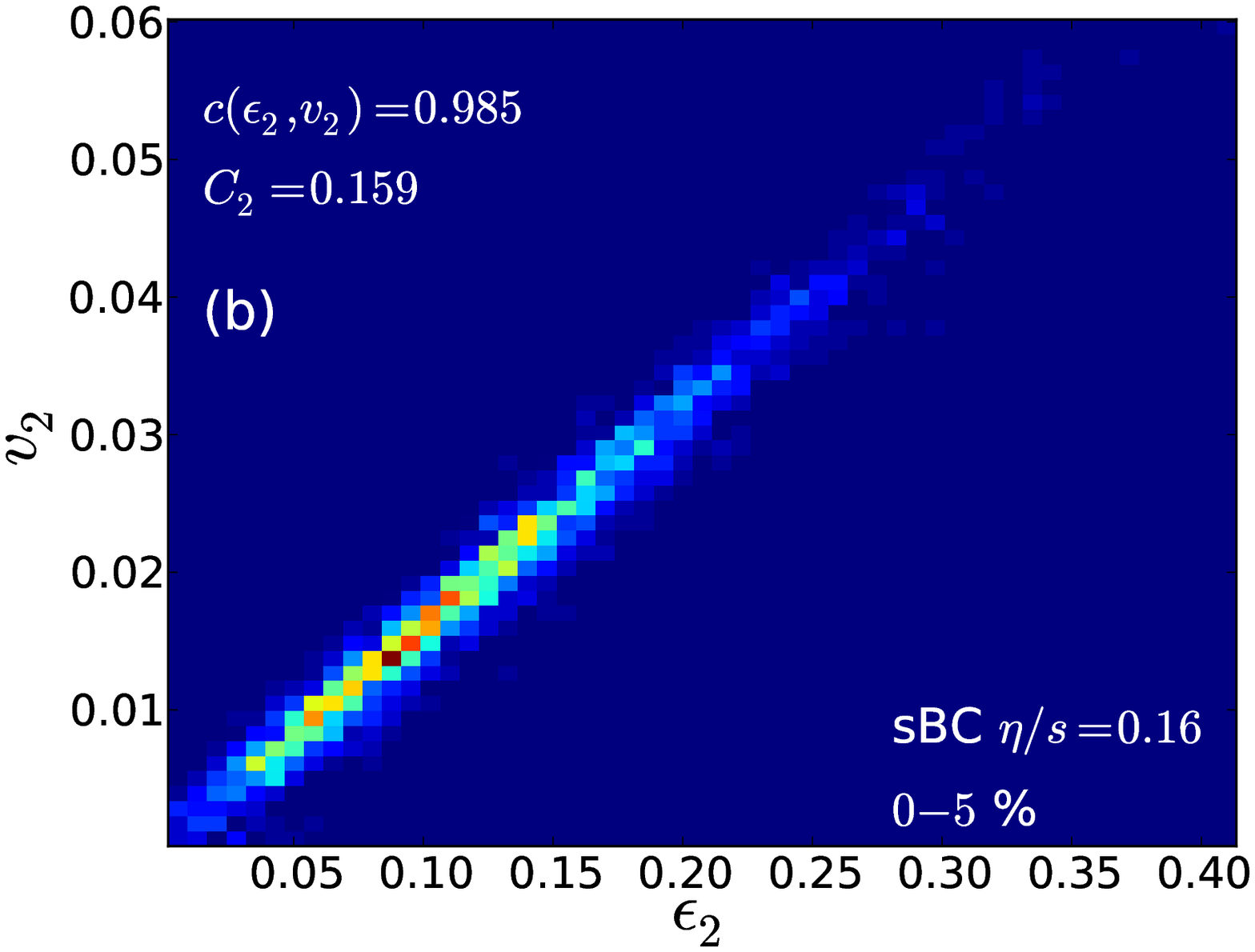} \epsfysize 4.4cm %
\epsfbox{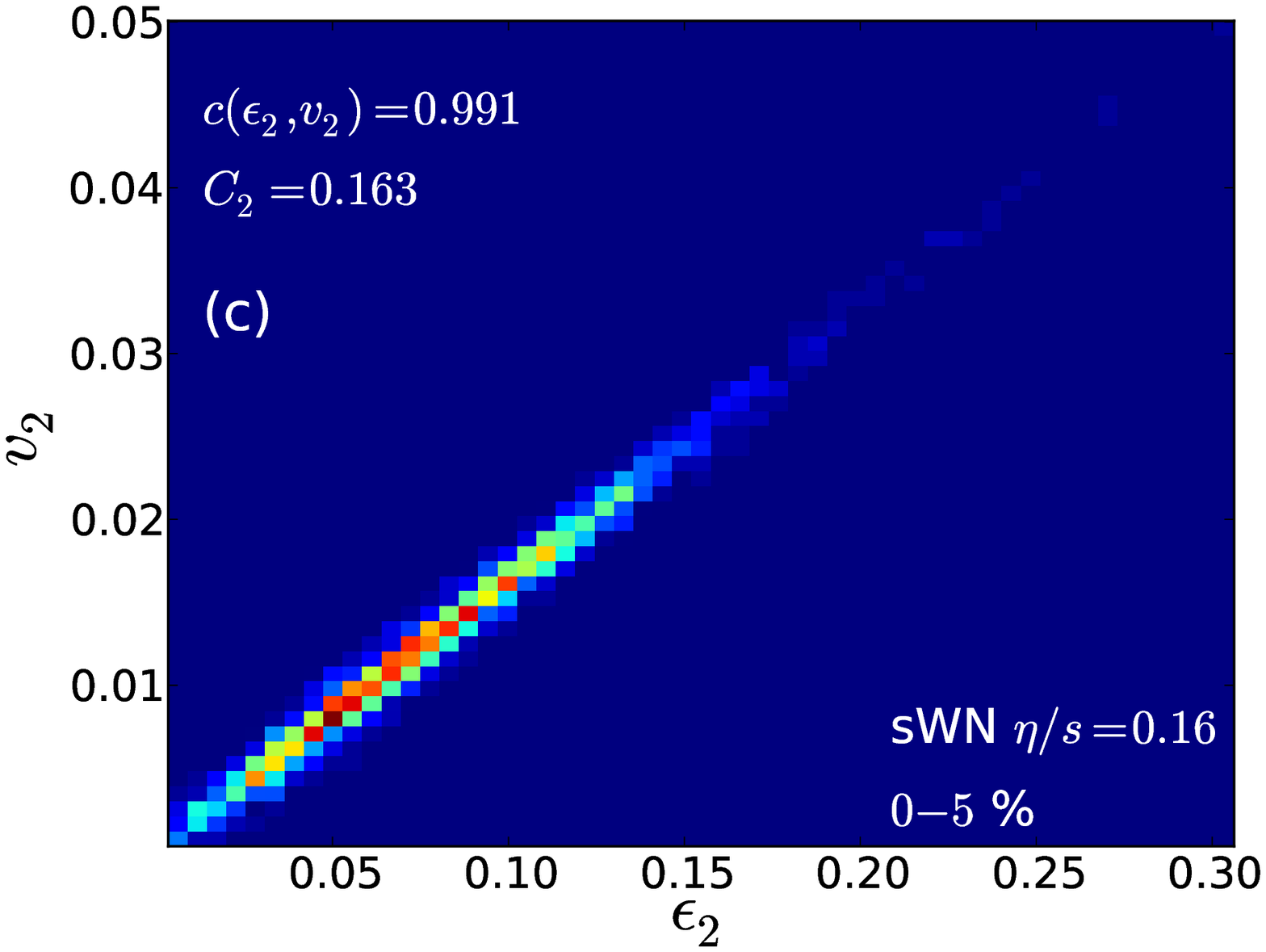} 
\caption{$\protect\epsilon_2$ and $v_2$ in the $0-5$\% centrality class
using different initializations and viscosities. a) sBC and $\protect\eta/s
= 0$, b) sBC and $\protect\eta/s = 0.16$ and c) sWN and $\protect\eta/s =
0.16$.}
\label{fig:e2v2_0_5}
\end{figure}
\begin{figure}[!h]
\hspace{-0.5cm} \epsfysize 4.4cm %
\epsfbox{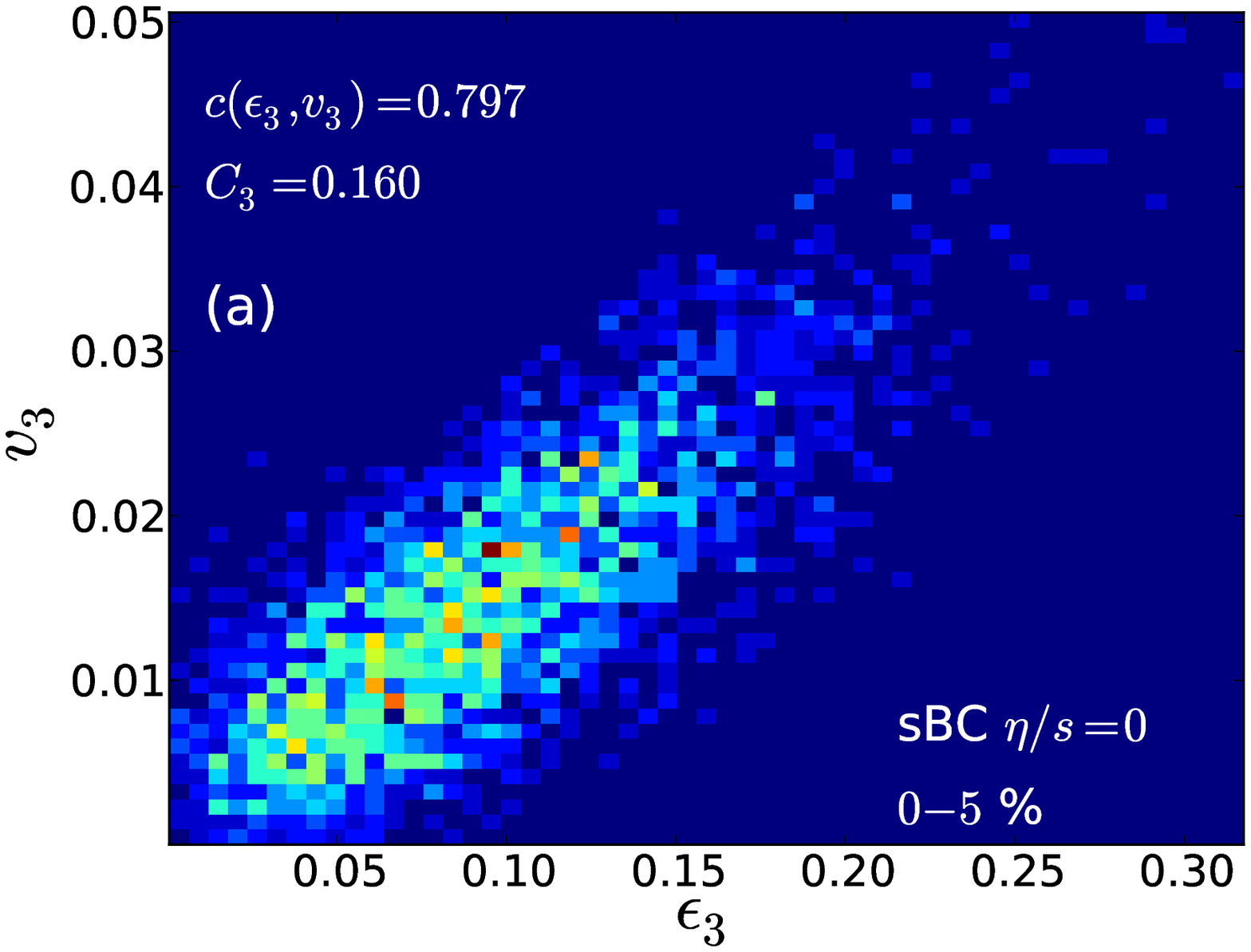} \hspace{0.0cm} \epsfysize %
4.4cm \epsfbox{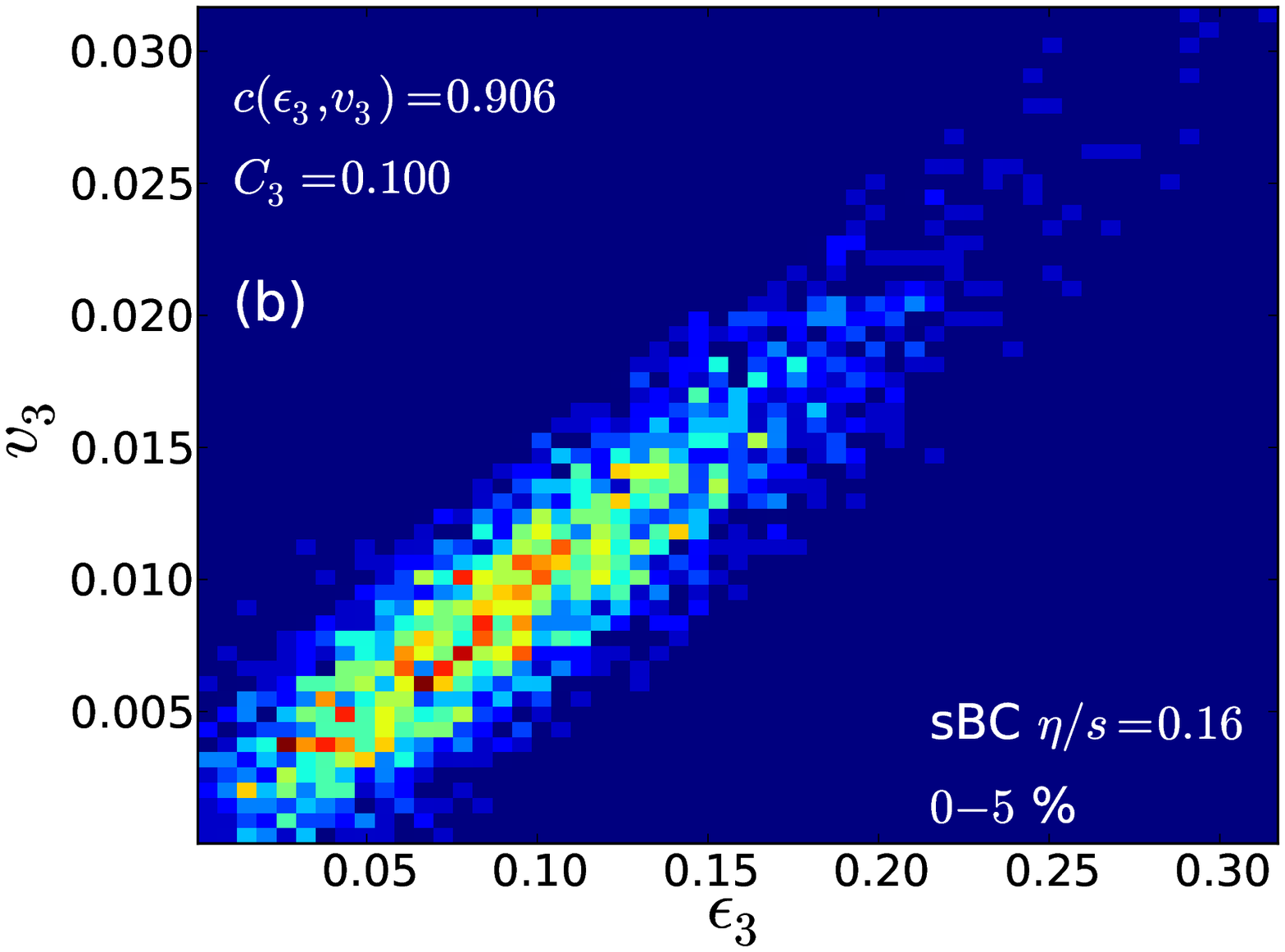} \epsfysize 4.4cm %
\epsfbox{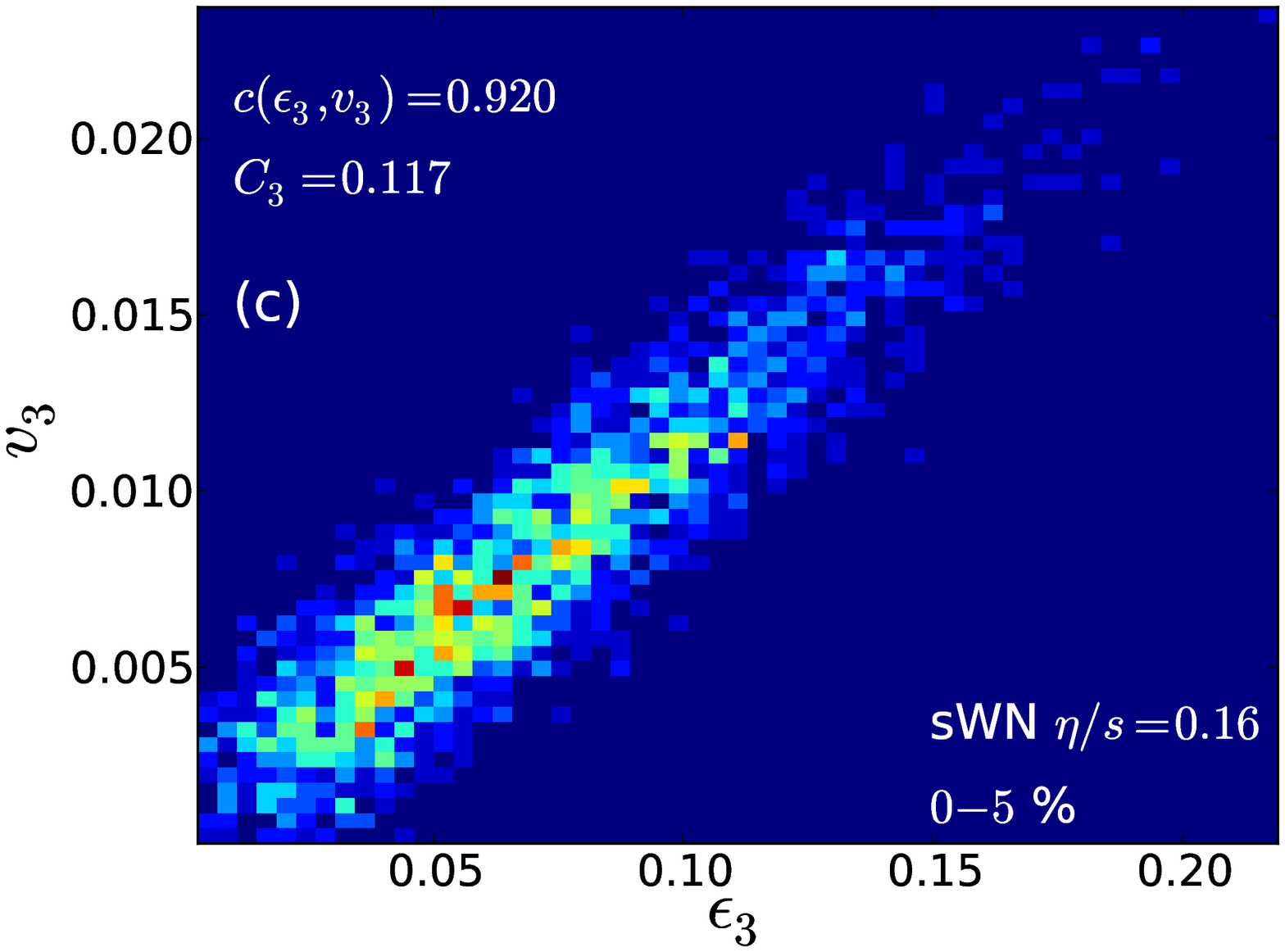} 
\caption{$\protect\epsilon_3$ and $v_3$ in the $0-5$\% centrality class
using different initializations and viscosities. a) sBC and $\protect\eta/s
= 0$, b) sBC and $\protect\eta/s = 0.16$ and c) sWN and $\protect\eta/s =
0.16$.}
\label{fig:e3v3_0_5}
\end{figure}
\begin{figure}[!h]
\hspace{-0.5cm} \epsfysize 4.4cm %
\epsfbox{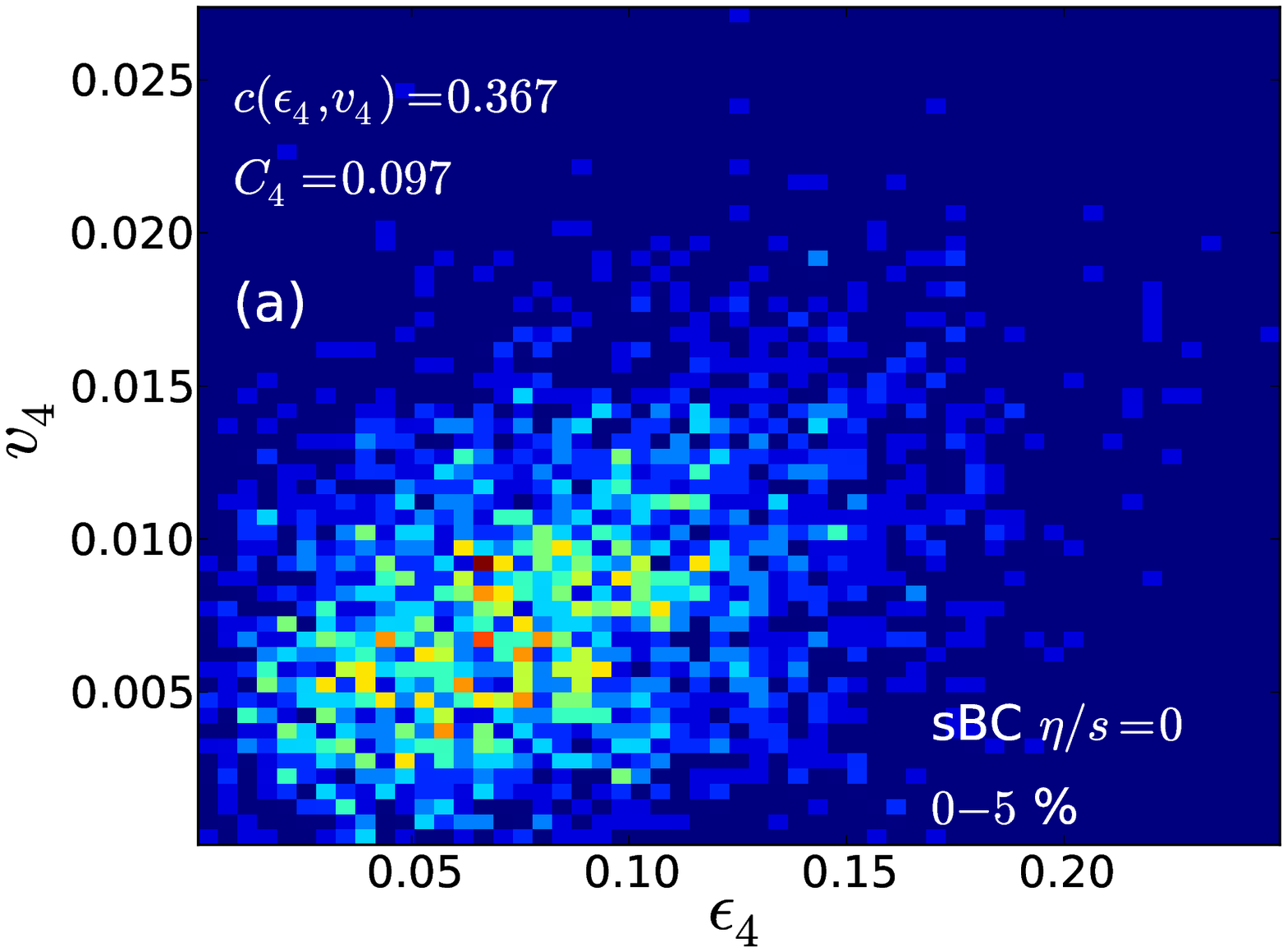} \hspace{0.0cm} \epsfysize %
4.4cm \epsfbox{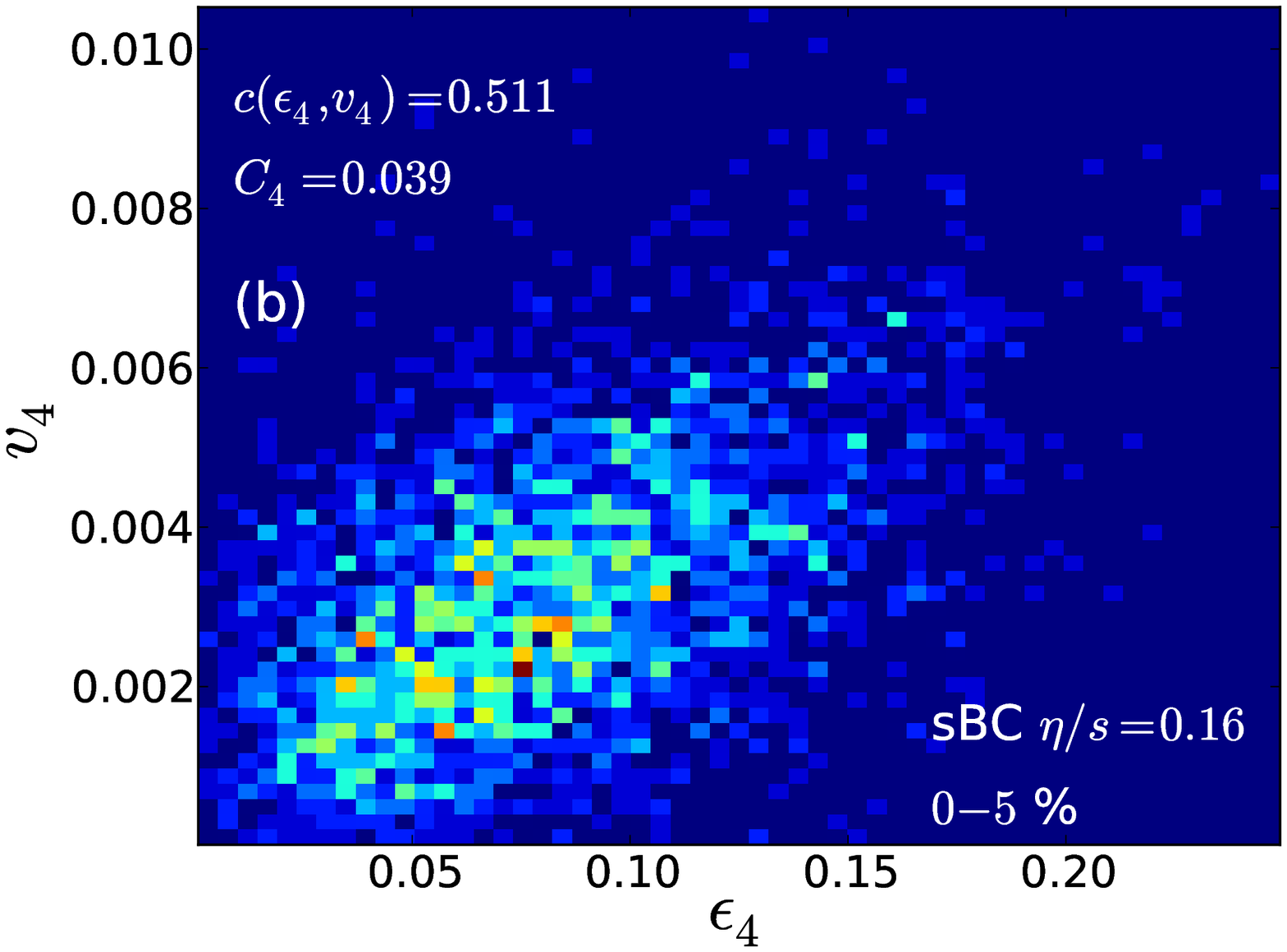} \epsfysize 4.4cm %
\epsfbox{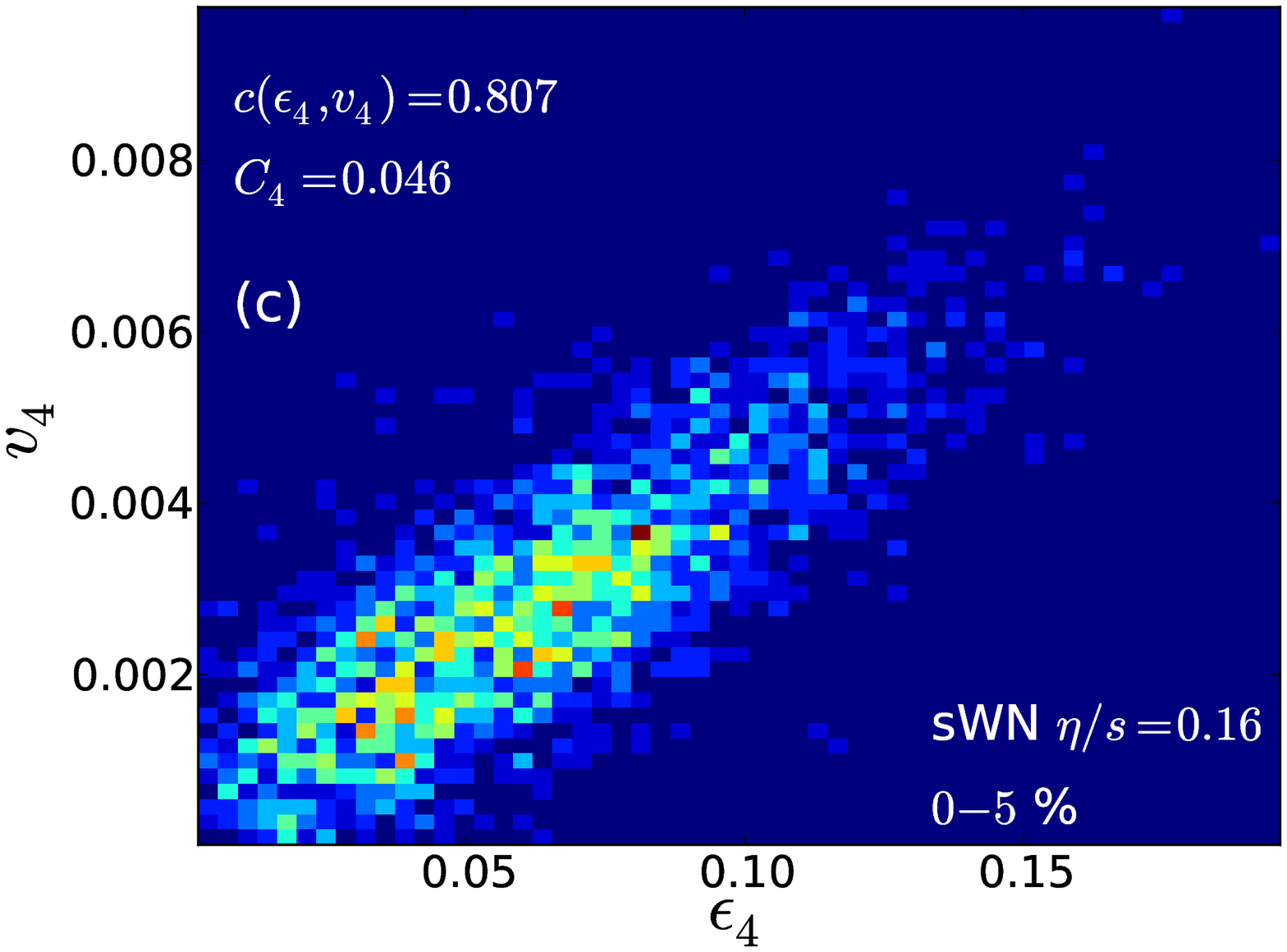} 
\caption{$\protect\epsilon_4$ and $v_4$ in the $0-5$\% centrality class
using different initializations and viscosities. a) sBC and $\protect\eta/s
= 0$, b) sBC and $\protect\eta/s = 0.16$ and c) sWN and $\protect\eta/s =
0.16$.}
\label{fig:e4v4_0_5}
\end{figure}

In Figs.~\ref{fig:e2v2_0_5} and \ref{fig:e3v3_0_5} we show the
2-dimensional histograms of $\epsilon_2$ and $v_2$ and
of $\epsilon_3$ and $v_3$,
respectively, in the $0-5$ \% centrality class. We plot the same cases
considered above: (a) sBC initialization with $\eta/s = 0$, (b) sBC
initialization with $\eta/s = 0.16$, and (c) sWN initialization with
$\eta/s = 0.16$. For $n=2$ and $n=3$ the linear correlation is
still valid. Also, the effect of shear viscosity and 
initialization on $C_n$ remain qualitatively the same.
On the other hand, in Fig.~\ref{fig:e4v4_0_5} the correlation
between $\epsilon_4$ and $v_4$ in central
collisions is drastically different from the correlation in the $20-30$ \%
centrality class. In the $0-5$ \% centrality class the
linear correlation coefficient $c(\epsilon_4, v_4)$ becomes much closer to $1
$ when compared to the peripheral case. It can be as large as $\sim 0.81$
obtained for the sWN initialization with $\eta/s = 0.16$. This behavior is
expected since in Ref.~\cite{Gardim:2011xv} it was shown that
$\epsilon_4$ becomes a better estimator for $v_4$ in central collisions.

\subsection{distributions of $v_n$}

\label{distributions}

So far the event-averaged values of $v_n$ have been extensively
studied. In order to observe what can be learned by looking at $v_n$
probability distributions, it is convenient to remove the average
from the distributions, and study the relative fluctuations using
the scaled variables
\begin{equation}
 \delta v_n = \frac{v_n - \langle v_n\rangle_{\mathrm{ev}}}
                   {\langle v_n\rangle_{\mathrm{ev}}},\qquad\mathrm{and}\qquad
 \delta \epsilon_n = \frac{\epsilon_n - \langle \epsilon_n\rangle_{\mathrm{ev}}}
                   {\langle \epsilon_n\rangle_{\mathrm{ev}}}.
\end{equation}
In this way changes in the probability distributions due to changes in
the average values are removed.

It was shown in the previous subsection that $v_n$ and $\epsilon_n$
have a strong linear correlation for $n=2$ and 3. As discussed in the
Appendix, if two variables are linearly correlated, and
$\langle d \rangle = 0$, the variances of the
relative distributions are equal. Since viscosity has only a small
effect on the correlations of $v_n$ and $\epsilon_n$, we expect that
$\sigma^2_{v_n} \approx \sigma^2_{\epsilon_n}$, independent of
viscosity. In such a case the information about the fluid response to
the initial geometry is contained in the coefficients $C_n$
controlling the average $\langle v_n\rangle_{\mathrm{ev}}$, while the
relative fluctuations of $v_n$ originate from the relative
fluctuations of $\epsilon_n$ and do not depend on viscosity of the fluid.

\begin{figure}[!h]
\hspace{-0.5cm} \epsfysize 4.5cm %
\epsfbox{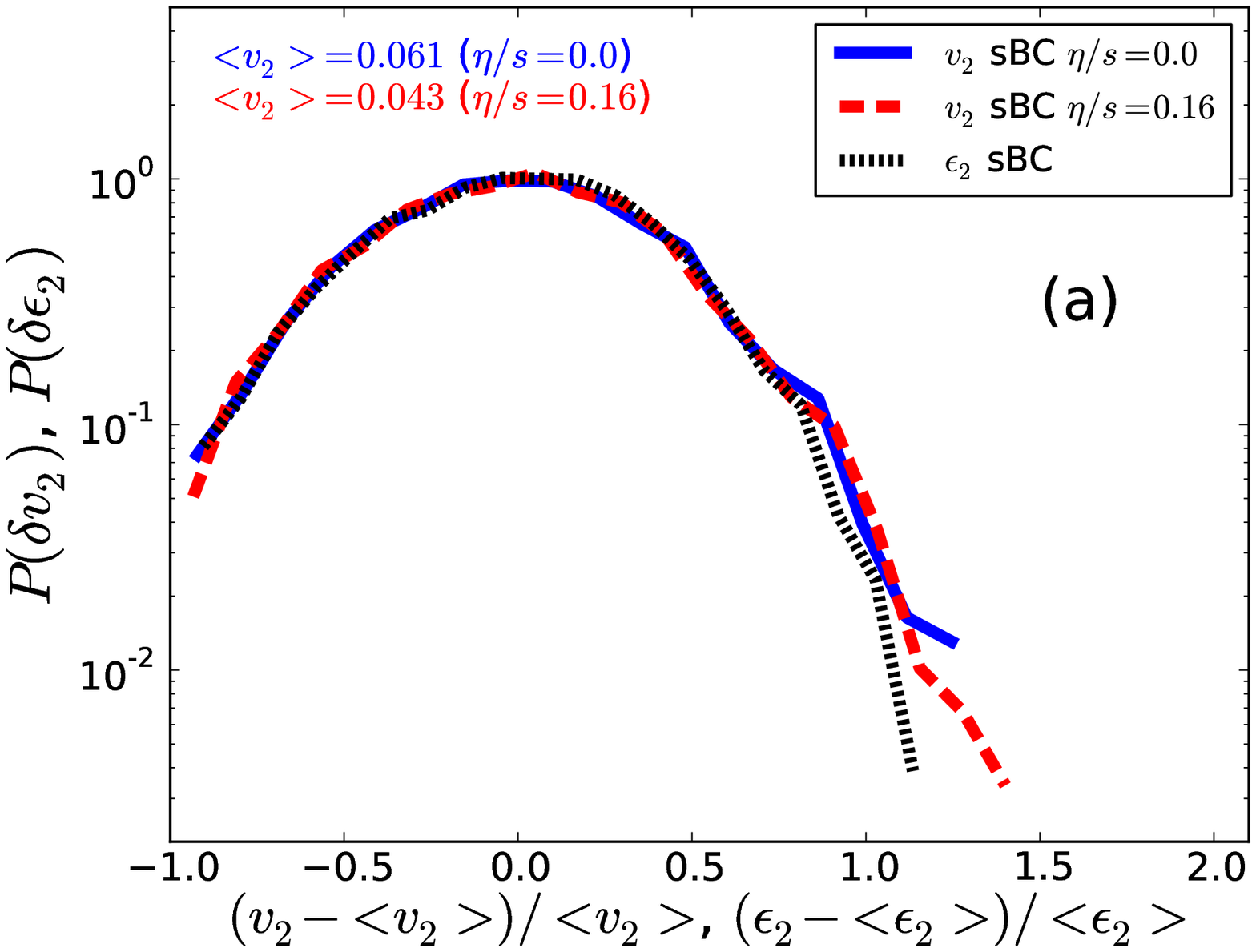} \hspace{0.0cm} \epsfysize %
4.5cm \epsfbox{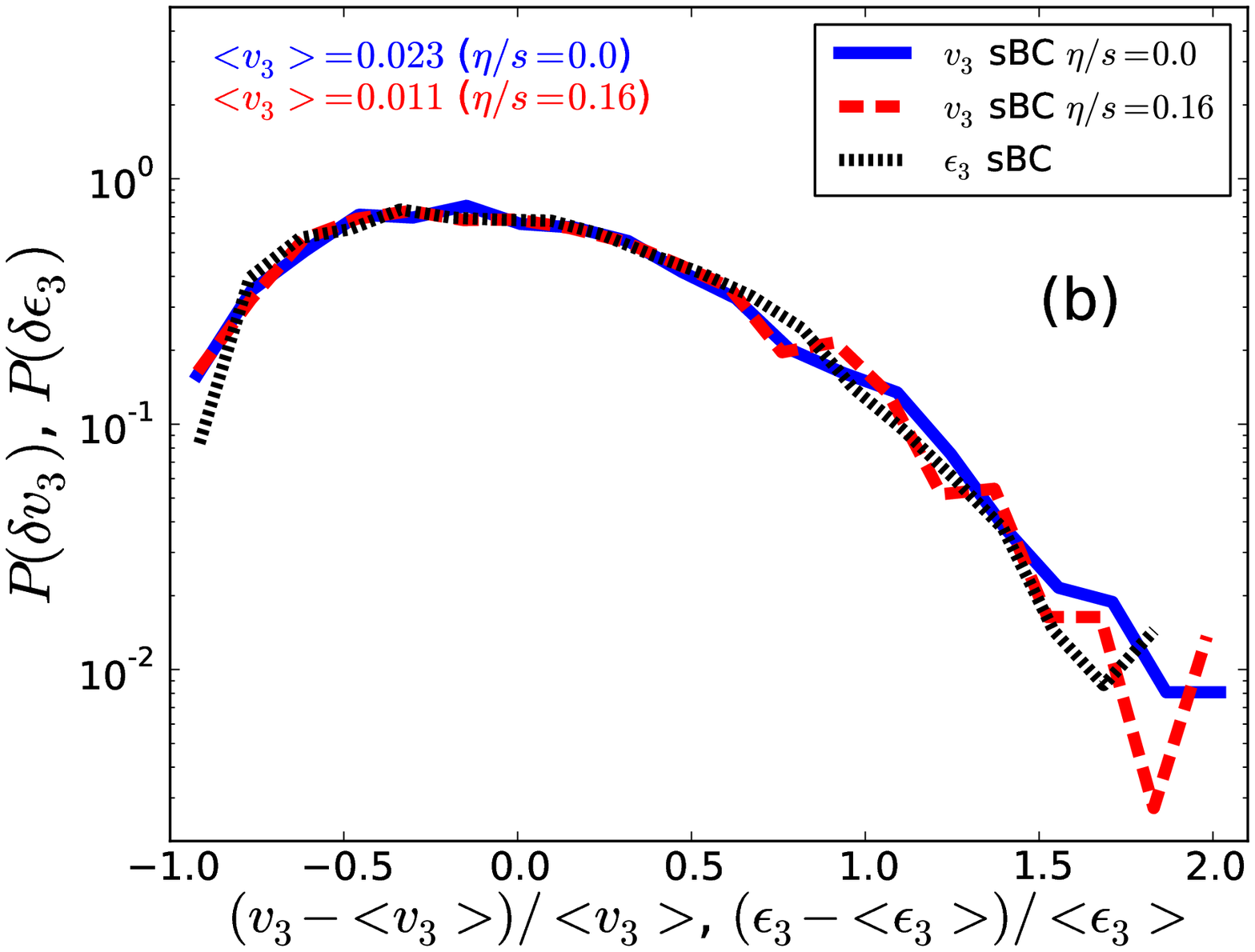} \epsfysize 4.5cm %
\epsfbox{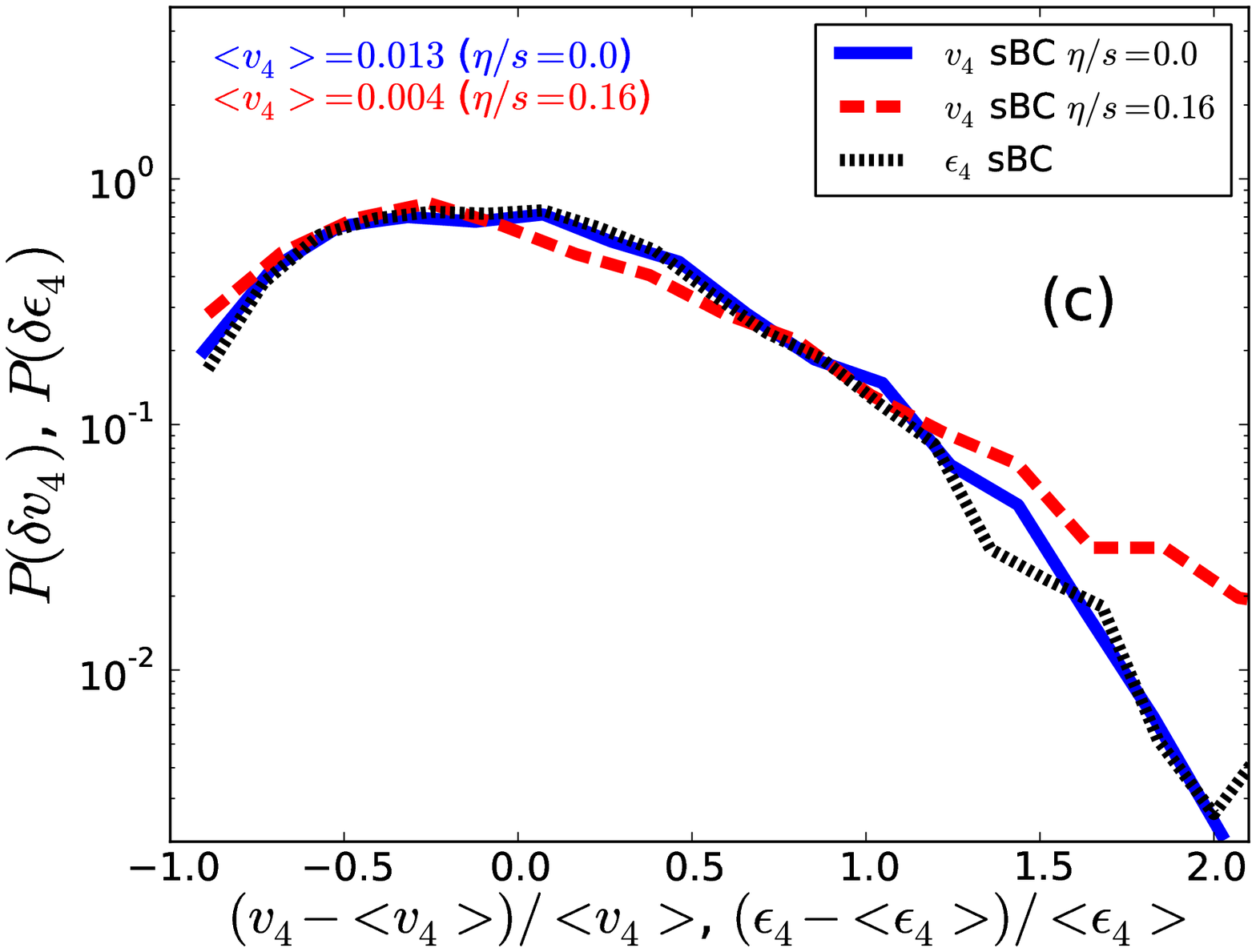} 
\caption{Probability distributions: a) $P(\protect\delta v_2)$ and $P(%
\protect\delta\protect\epsilon_2)$, b) $P(\protect\delta v_3)$ and $P(%
\protect\delta\protect\epsilon_3)$, and c) $P(\protect\delta v_4)$ and $P(%
\protect\delta\protect\epsilon_4)$, in the $20-30$ \% centrality class with
sBC initialization and two different values of $\protect\eta/s$, $\protect%
\eta/s = 0$ and $\protect\eta/s = 0.16$.}
\label{fig:dist_eta}
\end{figure}
\begin{figure}[!h]
\hspace{-0.5cm} \epsfysize 4.5cm %
\epsfbox{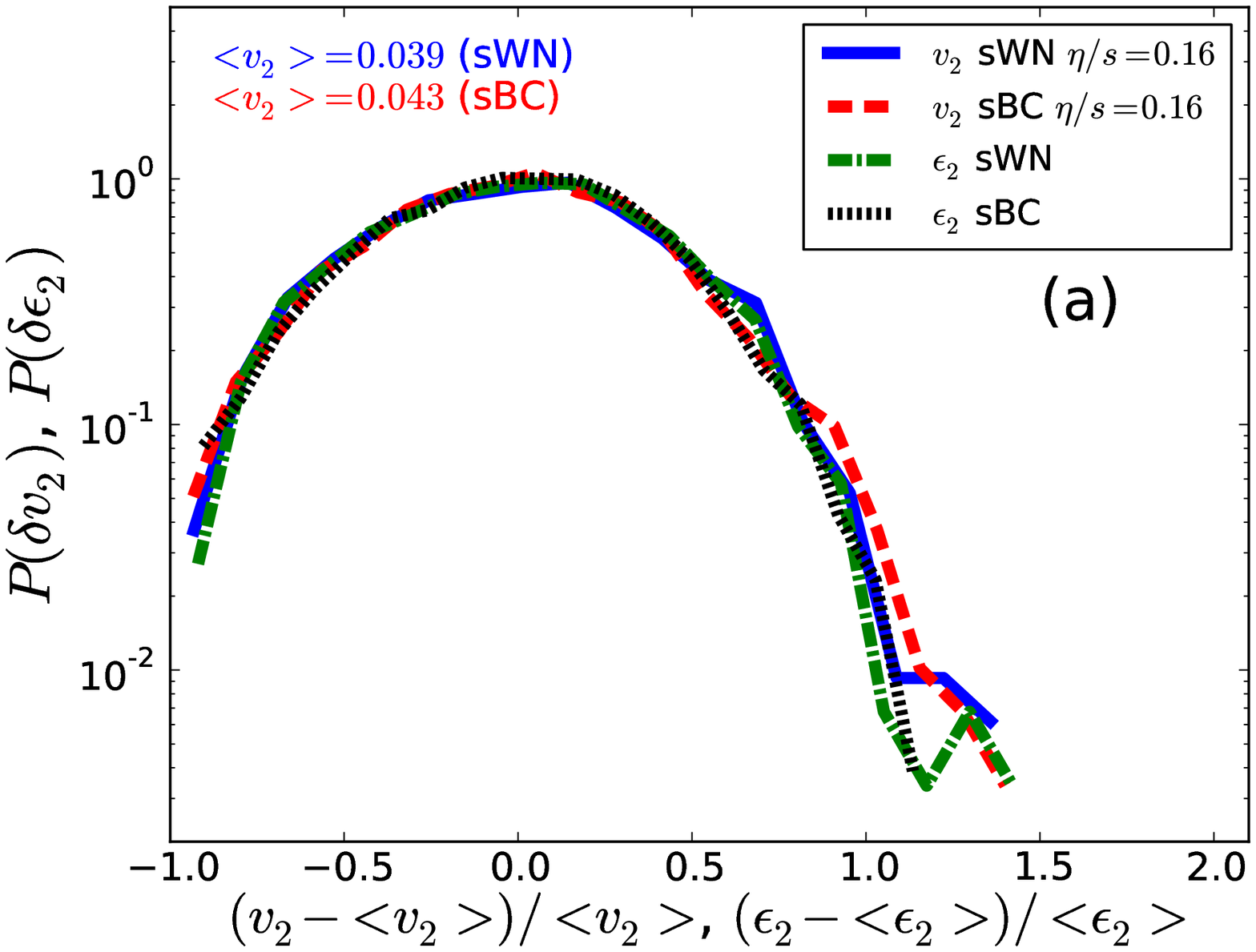} \hspace{0.0cm} \epsfysize %
4.5cm \epsfbox{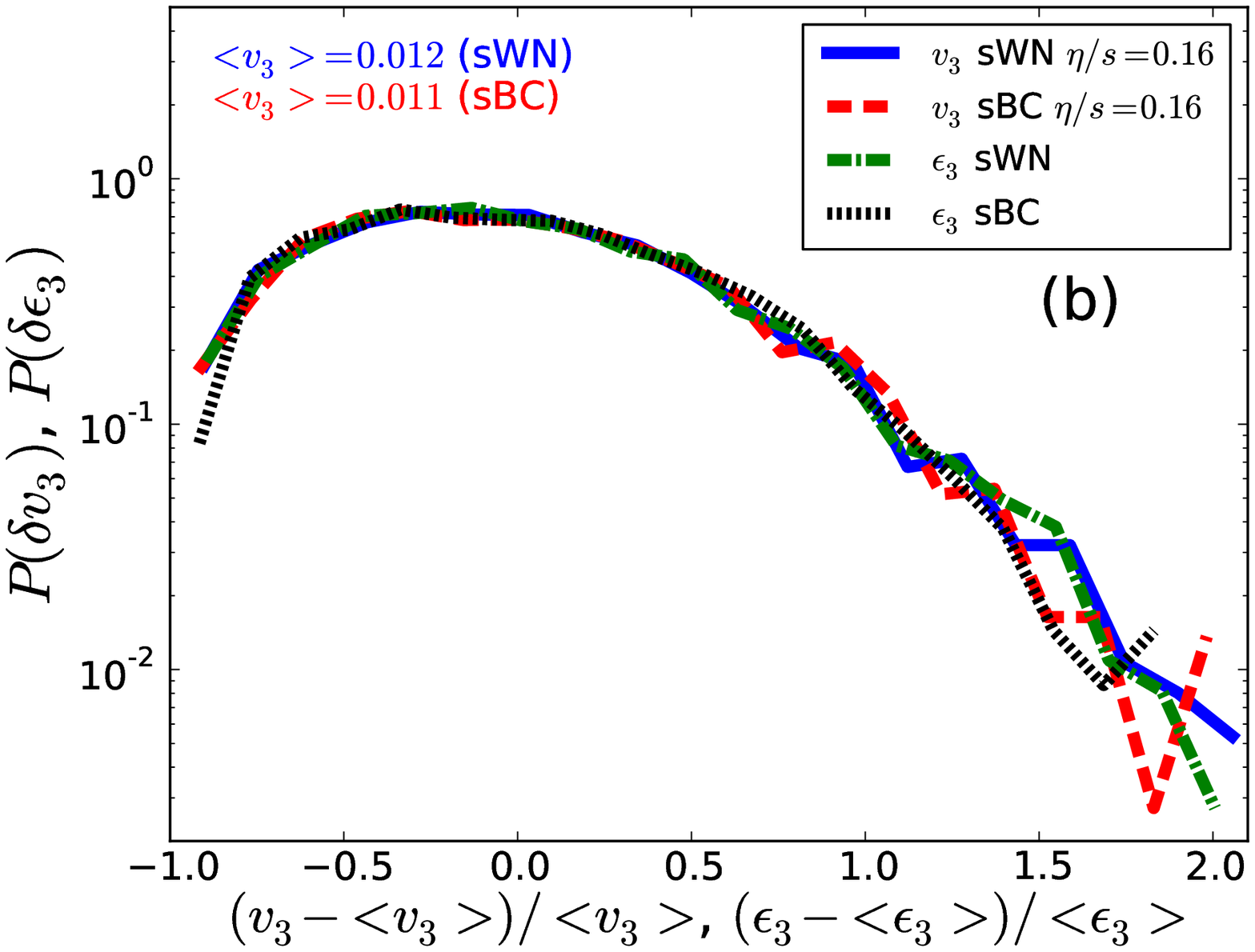} \epsfysize 4.5cm %
\epsfbox{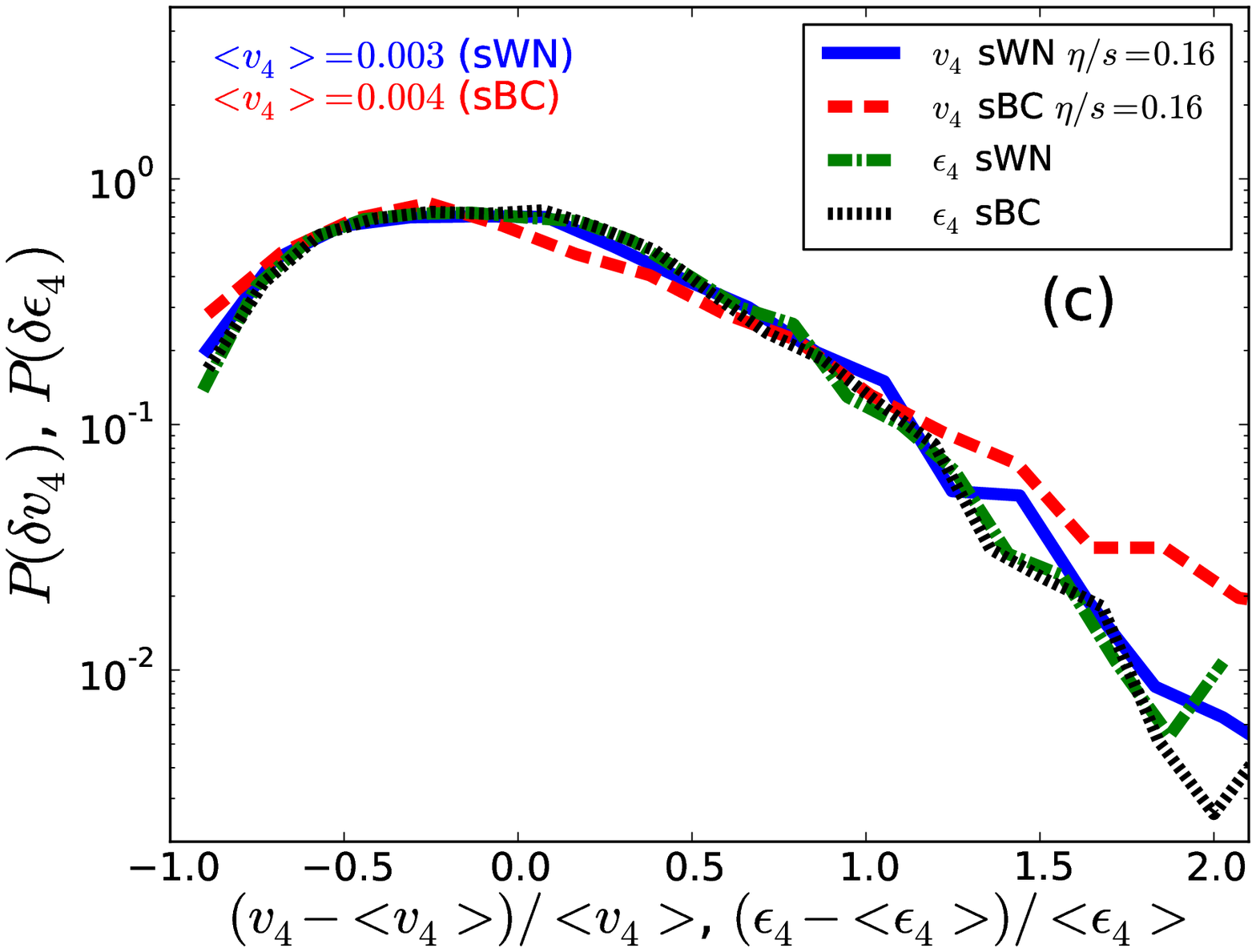} 
\caption{Probability distributions: a) $P(\protect\delta v_2)$ and $P(%
\protect\delta\protect\epsilon_2)$, b) $P(\protect\delta v_3)$ and $P(%
\protect\delta\protect\epsilon_3)$, and c) $P(\protect\delta v_4)$ and $P(%
\protect\delta\protect\epsilon_4)$, in the $20-30$ \% centrality class with $%
\protect\eta/s = 0.16$ and two different initial conditions, sBC and sWN.}
\label{fig:dist_ini}
\end{figure}

To test this assumption, and to see whether similarity extends
beyond variances, we plot the probability distributions 
$P(\delta v_{n})$ and $P(\delta \epsilon_{n})$ in $20-30$ \%
centrality class in Fig.~\ref{fig:dist_eta} using the sBC
initialization and two values of viscosity, $\eta/s = 0$ and
0.16. As seen in panel a) for $n=2$ and in panel b) for $n=3$, not
only the variances are similar, but the entire distributions are
almost identical. Fig.~\ref{fig:dist_eta}c depicts the relative
distributions for $n=4$, and surprisingly they are very similar even
if $v_4$ and $\epsilon_4$ are not linearly correlated. There are
deviations only at the tail of the distribution. As discussed in the
Appendix, lack of correlation leads to a large spread of possible
values of $v_4$ at given $\epsilon_4$ which tends to make $\delta
v_4$ distribution wider, but this effect can be canceled by other
terms. How these terms arise is an interesting question beyond the
scope of this paper.

To check whether the similarity of $P(\delta v_n)$ and
$P(\delta \epsilon_n)$ is only a coincidence based on the sBC
initialization, we show in Fig.~\ref{fig:dist_ini} the distributions
using both sBC and sWN
initialization, but using only one value of viscosity, 
$\eta/s = 0.16$. Again, panels a, b, and
c depict cases with $n=2$, 3, and 4, respectively. $P(\delta v_n)$
and $P(\delta \epsilon_n)$ are almost equal for both
initializations. The distributions $P(\delta \epsilon_n)$ are also similar
for both initializations, but this is because both Glauber-type
initializations give rise to the same relative fluctuations of
initial state anisotropies, see discussion in Ref.~\cite{Broniowski:2007ft}.

These results are valid in the $0-5$ \% centrality class as
well. We have also checked that $P(\delta v_n)$ is not sensitive
to the freeze-out temperature within interval $100 < T_\mathrm{fo} <
160$ MeV. Thus the
distribution of relative fluctuations 
of $v_{n}$
could be the ideal observable to study the fluctuations of the initial
geometry. If fluid dynamics provides 
a correct description of
heavy-ion collisions, $P(\delta v_n)$ is a direct measurement of the initial
state anisotropy fluctuations and can be compared directly to initial
condition models, such as Monte-Carlo Glauber or various implementations of
Color-Glass Condensate based initial conditions~\cite{AdrianSchenkeXXX}.

\subsection{$(v_n, v_m)$ linear correlations}

\label{Lcorrelations}

\begin{figure}[!h]
\hspace{-0.5cm} \epsfysize 4.5cm \epsfbox{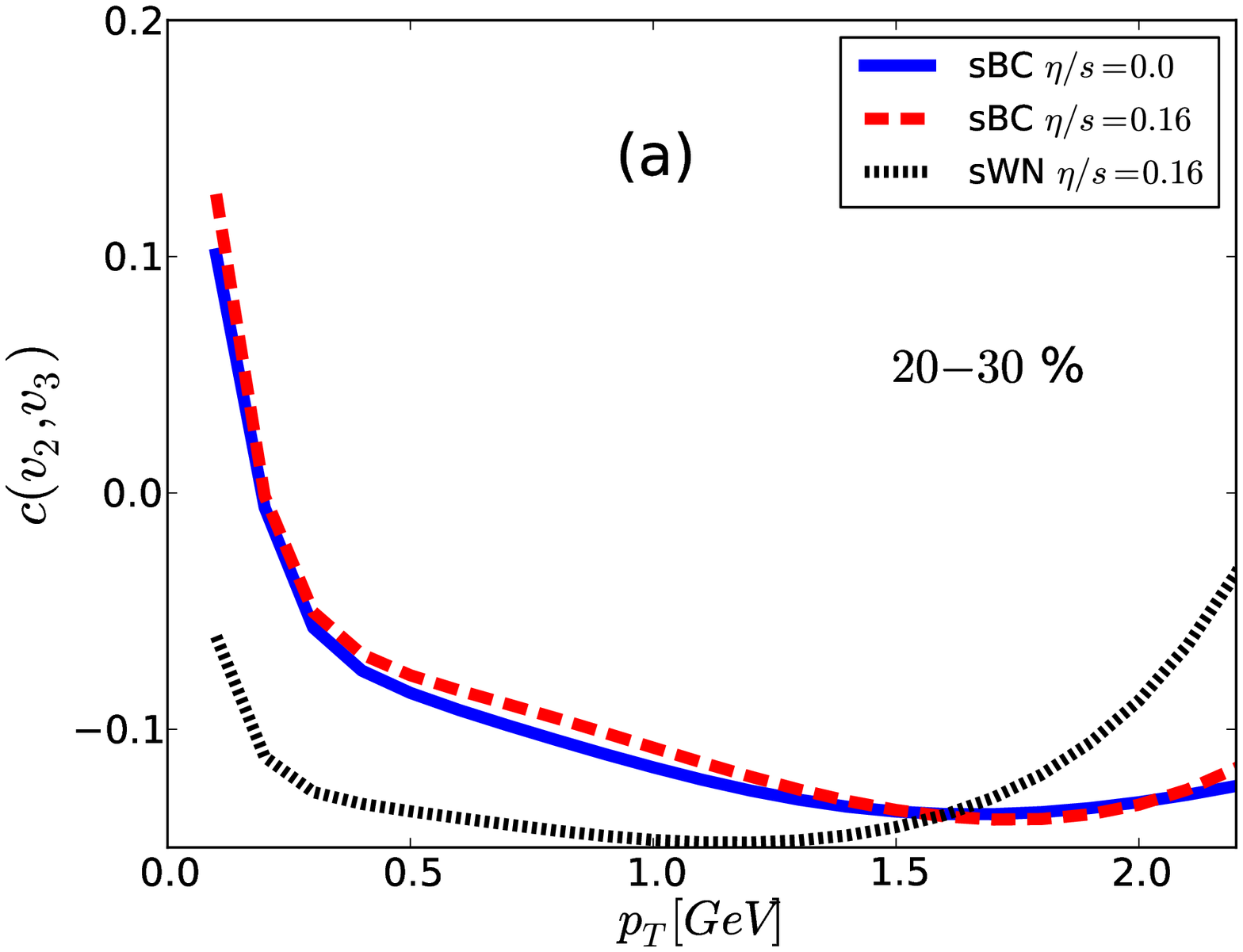} \hspace{%
0.0cm} \epsfysize 4.5cm \epsfbox{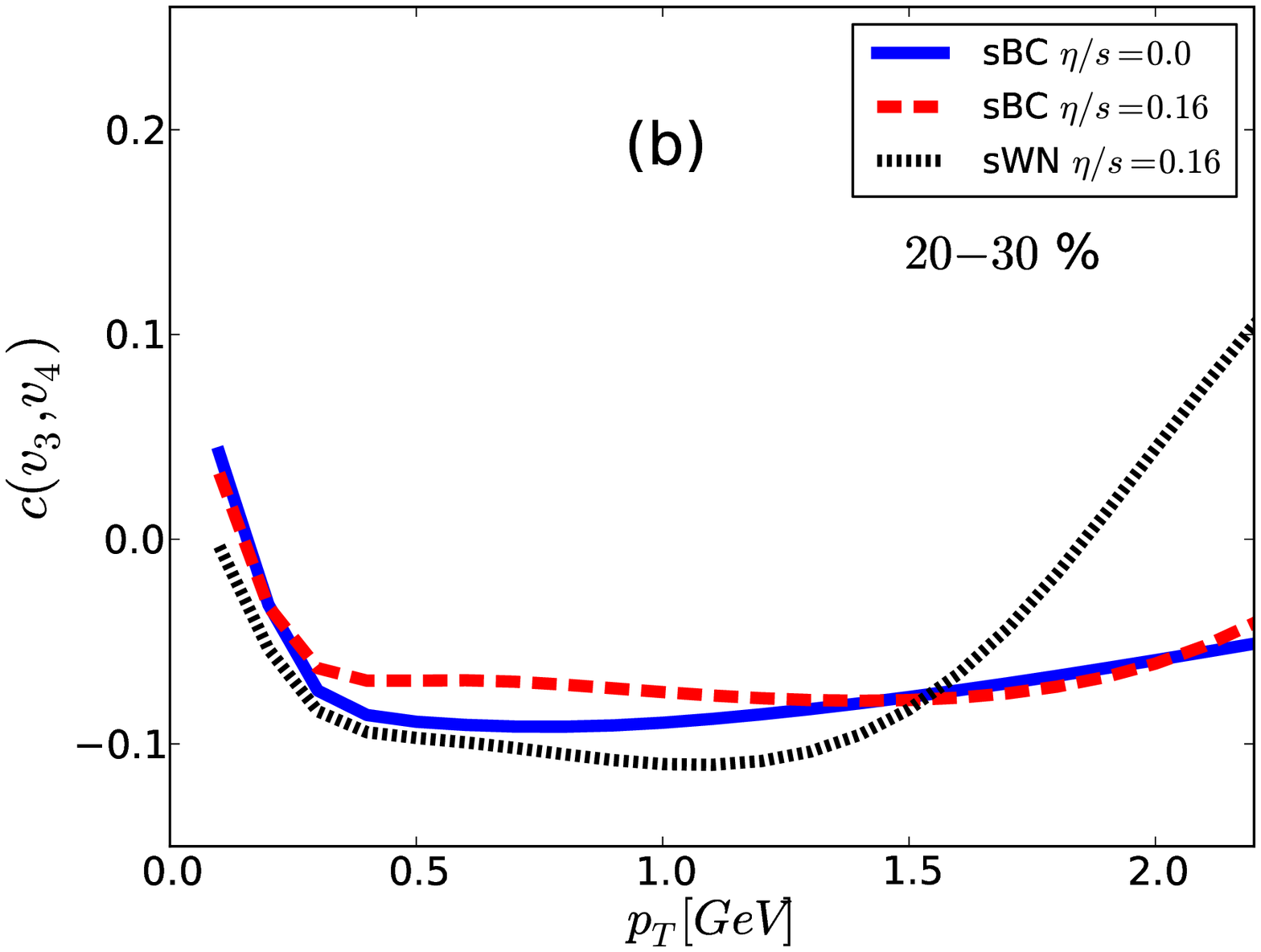} \epsfysize 4.5cm %
\epsfbox{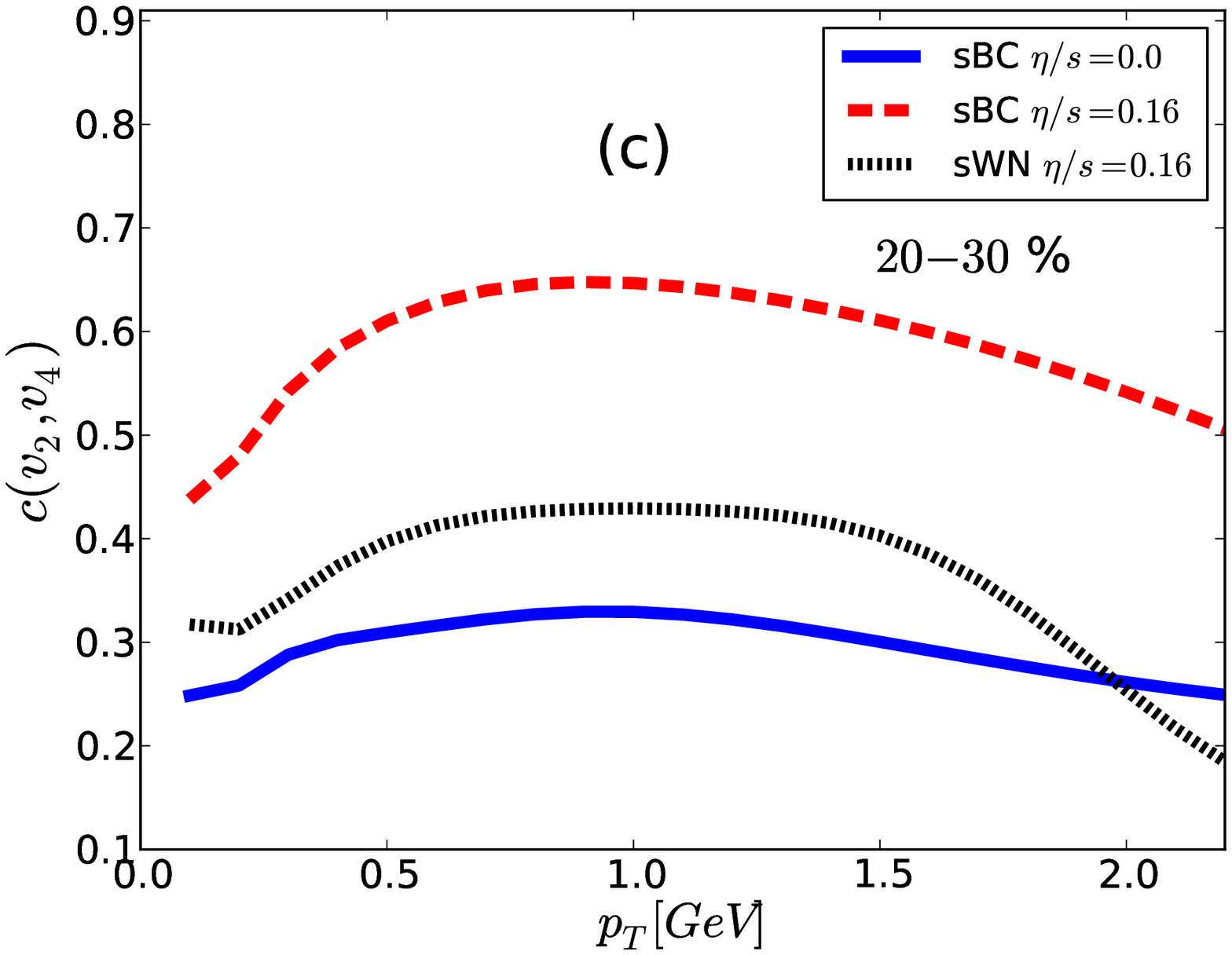} 
\caption{Correlations: a) $c(v_2,v_3)$, b) $c(v_3,v_4)$, and c) $c(v_2,v_4)$%
, as function of transverse momentum in the $20-30$ \% centrality class
using different initializations and viscosities.}
\label{fig:corr}
\end{figure}

\begin{table}[h]
\begin{center}
\begin{tabular}{|c|c|c|c|c|c|c|}
\hline
& $c(\epsilon_2, \epsilon_3)$ & $c(v_2, v_3)$ & $c(\epsilon_2, \epsilon_4)$
& $c(v_2, v_4)$ & $c(\epsilon_3, \epsilon_4)$ & $c(v_3, v_4)$ \\ \hline
sBC $\eta/s = 0.0$ & $-0.09$ & $-0.11$ & $0.26$ & $0.32$ & $-0.03$ & $-0.11$
\\ \hline
sBC $\eta/s = 0.16$ & $-0.09$ & $-0.11$ & $0.25$ & $0.63$ & $-0.03$ & $-0.09$
\\ \hline
sWN $\eta/s = 0.16$ & $-0.15$ & $-0.14$ & $-0.04$ & $0.42$ & $0.03$ & $-0.11$
\\ \hline
\end{tabular}%
\end{center}
\par
\vspace*{-0.1cm} \vspace*{-0.3cm}
\caption{{\protect\small Linear correlation coefficients for $p_T$%
-integrated $v_n$'s and $\protect\epsilon_n$'s in the $20-30$ \% centrality
class.}}
\label{tab:correlations}
\end{table}

We computed the linear correlation coefficients, $c\left( v_{2},v_{3}\right) 
$, $c\left( v_{2},v_{4}\right) $, and $c\left( v_{3},v_{4}\right) $ as a
function of the transverse momentum, $p_{T}$. We found that $c\left(
v_{2},v_{3}\right) \sim c\left( v_{3},v_{4}\right) \sim 0$ and, therefore,
are not \emph{linearly} correlated. These correlations are shown in Fig.~\ref%
{fig:corr}a and \ref{fig:corr}b. We also show the values of the correlation
coefficient between the integrated $v_n$'s and of the 
coefficients
$\epsilon_n$ in Table~\ref{tab:correlations}. It should be noted that, even
though both Glauber initializations used in this paper have the same
relative anisotropy fluctuations, their correlations differ.

Figure \ref{fig:corr}c shows the correlation coefficient between $v_{2}$ and 
$v_{4}$ as a function of $p_{T}$. As can be read off from the Figure, $%
c(v_{2},v_{4})$ depends strongly on $\eta /s$ and, consequently, is
sensitive to the properties of the QGP. Also, this correlation function is
strongly affected by the correlation between $\epsilon_{2}$ and $\epsilon_{4}
$ existing in the initial state, i.e., initial conditions with different $%
c(\epsilon_{2},\epsilon_{4})$ lead to very different $c(v_{2},v_{4})$, see
the dashed and dotted lines in Fig.~\ref{fig:corr}c. Overall, just like $%
v_{2}$ probes the fluid-dynamical response to an initial geometry
(characterized by $\epsilon_{2}$), $c(v_{2},v_{4})$ probes the
fluid-dynamical response to correlations in the initial geometry (in this
case characterized by $c(\epsilon_{2},\epsilon_{4})$).

In our investigations, $c(v_{2},v_{4})$ is the only correlation that was
sensitive to both the fluctuations of the initial condition and
the transport properties of the fluid. 
This correlation can thus be used as a further constraint to the
fluid-dynamical models applied to heavy-ion collisions.

\section{Conclusions}

\label{conclusions}

In this paper, we studied the relation between $v_{n}$ and $\epsilon _{n}$
in ultrarelativistic heavy-ion collisions using event-by-event fluid
dynamics. We confirmed that the second and third Fourier coefficients have a
strong linear correlation to the initial geometry of the collision. We
showed that while the event-average Fourier coefficients, $\left\langle
v_{n}\right\rangle _{\mathrm{ev}}$, $n=2$, $3$, and $4$, are sensitive to the
details of the fluid-dynamical evolution, their relative fluctuations, 
$\delta v_n =
\left( v_{n}-\left\langle v_{n}\right\rangle _{\mathrm{ev}}\right)/ 
\left\langle v_{n}\right\rangle _{\mathrm{ev}}$, are determined solely by
the fluctuations of the corresponding initial state anisotropy
coefficients, with basically no sensitivity to the viscosity of the
fluid. This makes the distribution of $\delta v_n$ a direct probe of
the initial condition of a heavy-ion collision, providing a direct and
clean measurement of the distribution of the relative fluctuations of
the initial anisotropy, \emph{i.e.},
\begin{equation}
 P\left(\delta v_n\right) \simeq P\left(\delta \epsilon_n\right) 
 ,\text{ \ }n=2,3,4\text{.}
\end{equation}
Surprisingly, this relation was shown to be true even for the relative
fluctuations of $v_{4}$, even though $v_{4}$ itself is not linearly
correlated to $\epsilon_{4}$.

Furthermore we found that the linear correlations between the flow harmonics 
$v_n$ are not solely dominated by the initial conditions. Especially the
correlation function $c\left(v_{2},v_{4}\right)$ is sensitive to the
viscosity of the fluid providing an additional constraint for the model
when one tries to
extract the viscosity coefficient from the data.

\section*{Acknowledgements}

The authors thank D.H.\ Rischke and J.\ Jia for discussions.
The work of H.N.\ was supported by Academy of Finland, Project No.~133005,
G.S.D.\ by Helmholz Research School H-QM, H.H.\ by the ExtreMe Matter
Institute (EMMI), and P.H.\ by BMBF under contract no.\ 06FY9092. We
acknowledge CSC --- IT Center for Science in Espoo, Finland, for the
allocation of computational resources.

\appendix*
\section{Relative widths of correlated distributions}

A relation between variables $\epsilon$ and $v$ (we omit the indexes
$n$ here) can be written as
\begin{equation}
 v = k\epsilon + d.
\end{equation}
If we define 
\begin{equation}
 k = \frac{\sigma_v}{\sigma_\epsilon}c(\epsilon,v),
\end{equation}
where $c(\epsilon,v)$ is the linear correlator (Eq.~(\ref{pearson})),
$\epsilon$ and $d$ are not linearly correlated, $c(\e,d) = 0$. It also
follows that
\begin{equation}
 \sigma^2_d = \sigma^2_v\left(1-c(\e,v)^2\right),
\end{equation}
\emph{i.e.} the stronger the linear (anti)correlation between $\e$ and
$d$, the narrower the distribution of $d$.

For the distributions of the scaled variables, 
$\delta x = (x-\langle x \rangle)/\langle x \rangle$, it holds that 
$\sigma_{\delta x} = \sigma_x/\langle x\rangle$, and that 
$\langle \delta x \rangle = 0$. The variance of $\delta v$ can now be
written as
\begin{eqnarray}
 \label{equal}
 \sigma^2_{\delta v} 
  & = & \frac{1}{\langle v\rangle^2}  \nonumber
         \left[k^2\sigma^2_\e + 2k\sigma_\e\sigma_dc(\e,d) + \sigma^2_d\right] \\
  & = & \frac{\sigma^2_\e}{\langle \e\rangle^2}
         \left[\frac{\langle \epsilon\rangle^2}{\langle v\rangle^2} k^2
              +\frac{\langle \epsilon\rangle^2}{\langle v\rangle^2}
               \frac{\sigma^2_d}{\sigma^2_\e}\right] \nonumber                \\
  & = & \sigma^2_{\delta \epsilon}
         \left[\frac{1}{\left(1+\frac{\langle d\rangle}{k\langle\epsilon\rangle}
                        \right)^2}
              +\frac{\langle \epsilon\rangle^2}{\langle v\rangle^2}
               \frac{\sigma^2_d}{\sigma^2_\e}\right].
\end{eqnarray}
As mentioned, if $c(\epsilon,v) \approx 1$, $\sigma_d \approx 0$, and
the last term in Eq.~(\ref{equal}) is negligible. If also
$\langle d\rangle \approx 0$, 
$\sigma_{\delta v} \approx \sigma_{\delta \epsilon}$. Roughly speaking
the requirement $\langle d\rangle \approx 0$ means that when 
$\epsilon\rightarrow 0$, $\langle v\rangle \approx 0$---a requirement
our distributions with $n=2$ and 3 fulfill as seen in
Figs.~\ref{fig:e2v2_20_30}, \ref{fig:e3v3_20_30}, \ref{fig:e2v2_0_5}, 
and \ref{fig:e3v3_0_5}. On the other hand, the correlation is weak for
$n=4$ in the $20-30$ \% centrality class. In that case the $d$ distribution
is wide, but $\langle d\rangle \neq 0$ as well, and the two terms in
Eq.~(\ref{equal}) sum to approximately one, and the distributions of
$\delta \epsilon$ and $\delta v$ are equal even in that case.

\end{document}